\documentclass[epsfig,psfig,aps,twocolumn,prb,showpacs]{revtex4-1}
\usepackage[colorlinks=true,citecolor=red,urlcolor=blue]{hyperref}
\usepackage{amsfonts}
\usepackage{graphicx}
\usepackage{epsfig}
\newcommand{\sign}{\mathrm{sign}} 
\begin{document}
%\draft
\title
{
Strain tuned topology in the Haldane and the modified Haldane models \\ 
}
\author
{
Marwa Manna\"{i} and Sonia Haddad$^{\ast}$
} 
\affiliation{
Laboratoire de Physique de la Mati\`ere Condens\'ee, D\'epartement de Physique,
Facult\'e des Sciences de Tunis, Universit\'e Tunis El Manar, Campus Universitaire 1060 Tunis, Tunisia
}
%\maketitle
%\date{\today}
%
%---------Abstract---------
%

\begin{abstract}
We study the interplay between a uniaxial strain and the topology of the Haldane and the modified Haldane models which, respectively, exhibit chiral and antichiral edge modes.
The latter were, recently, predicted by Colom\'es and Franz (Phys. Rev. Lett. {\bf 120}, 086603 (2018)) and expected to take place in the transition metal dichalcogenides.
Using the continuum approximation and a tight-binding approach, we investigate the effect of the strain on the topological phases and the corresponding edge modes. We show that the strain could induce transitions between topological phases with opposite Chern numbers or tune a topological phase into a trivial one. As a consequence, the dispersions of the chiral and antichiral edge modes are found to be strain dependent. The strain may reverse the direction of propagation of these modes and eventually destroy them. This effect may be used for strain-tunable edge currents in topological insulators and two-dimensional transition metal dichalcogenides. 
\end{abstract}

%\keywords{...}
\maketitle

\section{Introduction}
Topological insulators were first introduced by Haldane in his seminal paper \cite{Haldane} where he showed that the Hall conductance may be quantized in the absence of an external magnetic field, which is known as the Quantum Anomalous Hall (QAH) effect \cite{RevQAH1,RevQAH2,RevQAH3}. The latter arises as an intrinsic property of the electronic band structure. Haldane proposed a two-band spinless fermion model on a honeycomb lattice with local magnetic fluxes, breaking the time reversal symmetry (TRS), and arranged in a geometry resulting in a zero net flux per unit cell. The key parameters of the model are the Semenoff \cite{Semenoff} mass $M$ and the complex second nearest neighbor hopping integrals $t_2e^{\pm i\Phi}$. The sublattice potential $\pm M$, describing the masses of the two atoms, forming the lattice, is responsible of the inversion symmetry breaking while the complex hopping terms break the TRS due to the phase $\Phi$ acquired in the presence of the local magnetic fluxes.
By tuning the values of $M$ and $\Phi$, the ground state of the system undergoes transitions between phases with different topology characterized by a topological invariant, known as the first Chern number $C$ \cite{Thouless}. A Trivial, or band, insulating state corresponds to $C=0$ while a topological, or a Chern, insulator is described by a nonvanishing Chern number, which is $C=\pm 1$ in the case of the Haldane model (HM). Large Chern numbers are expected by taking into account distant neighbor hopping terms \cite{Fred13}.\

Besides the non-vanishing Chern number, the topological signature of a phase is marked by the presence of chiral edge states crossing the bulk gap of the band structure of a finite size system, as those found in the quantum Hall effect \cite{Halperin}. 
These edge states are at the origin of the substantial interest devoted to the QAH effect considered as a suitable candidate to pave the way for dissipationless electronic applications in the absence of a magnetic field \cite{RevQAH3}.\

Regarding the difficulty to fulfill the local magnetic flux requirements, the HM is, as stated by Haldane, {\it unlikely to be realized} in condensed matter \cite{Haldane}. The first realization of the HM was achieved with cold atoms in a shaken optical lattice \cite{Esslinger14}. The Fe-based honeycomb ferromagnetic insulators \cite{Kee} and transition-metal pnictides \cite{Huang18} are also expected to be described by the HM. \
The experimental realization of the QAH effect predicted by Haldane, became possible only after the prediction of the quantum spin Hall (QSH) effect, resulting from the generalization of the HM to the spinfull system with TRS invariance \cite{Kane}. This effect led to the discovery of the topological insulators considered as one of the hottest topics of interest in condensed matter physics \cite{Hasan,Zhang10,TI10,TI16,TI19}.
Several observations of QAH effect have been reported in magnetic topological insulators \cite{Chang13,Tokura14,Chang15}.\

To understand the fundamental aspects of the QAH effect it is necessary to uncover its dependence on the different external and intrinsic factors such as doping\cite{HM-real,doping}, disorder \cite{disorder,castro-disorder,Moessner}, temperature\cite{temperature}, interaction \cite{Roy,Pesin}, magnetic-electric fields\cite{field}, material thickness \cite{thick}, mechanical strain \cite{Roy,Zhu13,Chen13,Ghaemi,Castro17}. In particular, the latter is found to be a useful tool to tune the electronic band structure of graphene \cite{gr-strain} and topological insulators \cite{Roy,Kane11,Molenkamp11,Pesin,Liu14,Guassi,Peeters,Aramberri17,Mutch}.
Recently, the effect of a nonuniform strain on ribbons described by the HM was investigated within the tight-binding (TB) approach \cite{Castro17}. The authors showed that nonuniform strain does not affect the topological phases and the dispersion of the corresponding edge states.\

However, strain is found to be a substantial parameter to tune the properties of topological insulators \cite{Roy,Kane11,Molenkamp11,Pesin,Liu14,Guassi,Peeters,Aramberri17} and to induce helical edge states in armchair graphene nanoribbon \cite{Liu17}. Contrary to the chiral edge modes, occurring in systems with broken TRS, the helical edge states appear in systems where TRS is preserved, as in QSH effect \cite{Hasan}. As the chiral modes of the spinless QAH effect, the helical edge states propagate in opposite direction for a given spin.\newline
Recently, antichiral edge modes were proposed to occur in 2D semi-metal \cite{Colomes}, where co-propagating modes appear at the parallel edges of the system, and are counter-balanced by gapless bulk states. These edge states can be obtained in zigzag graphene nanoribbon described by the so-called modified Haldane model (mHM) where the Dirac points are offset in energy by a term $\pm 3\sqrt{3}t_2\sin \Phi$ \cite{mHM,Colomes}.\

In the present work, we raise the issue regarding the robustness of the topological phases of the HM and the mHM against a uniform uniaxial strain. We particularly, ask the following questions: Is it possible to tune, by the strain, the topology of these models? Could the direction of the propagation of chiral and antichiral edge modes be controlled by strain?

We first study the behavior of the Haldane phase diagram under a uniaxial strain using the continuum limit approximation and, then we derive, within a TB approach, the strain dependence of the edge states of a zigzag nanoribbon described by the HM.
In a second part, we consider the effect of the strain on the antichiral edge modes of the mHM.

The main results of this paper can be summarized as follows: (i) Contrary to a nonuniform deformation, uniaxial strain could destroy a topological phase and tune it to a trivial insulating state. (ii) By adjusting the strain amplitude, the system can be driven from a topological phase to another with opposite Chern number, which means that the strain may act on the edge current. (iii) At a tensile strain of $50\%$, transition between topological phases with opposite Chern numbers occurs on a line boundary and not only at the point $(M=0,\Phi=0)$ as found in the undeformed HM. This feature could not be realized in real crystals regarding the huge required strain value. However it could be observed in optical lattices of cold atoms \cite{Esslingermerge}. (iv) The antichiral edge modes, of the mHM, are strain dependent with a switchable energy dispersion. Such effect, which may lead to a strain-tunable edge currents, could be realized in two-dimensional (2D) transition metal dichalcogenides, as $\mathrm{WSe_2}$ showing edge states reminiscent of those of mHM \cite{Colomes}.\

The paper is organized as follows. In Sec. II, we describe the HM under uniaxial strain for an infinite honeycomb lattice, and then derive the strain dependence of the corresponding Chern number in the continuum limit. We then discuss the behavior of the phase diagram under the strain. In Sec. III, we consider, within the TB model, the effect of the strain on a zigzag nanoribbon described by the HM in the presence of a uniaxial strain applied along the armchair direction. We will focus on the behavior of the edge states as a function of the deformation. In Sec. IV, we discuss a zigzag graphene nanoribbon described by mHM under a uniaxial strain. We numerically determine, within the tight binding approach, the strain dependence of the corresponding antichiral edge mode.
Sec. V is devoted to the concluding remarks.

\section{Haldane model under uniaxial strain}
\subsection{Electronic Hamiltonian}

\begin{figure}[hpbt]
\includegraphics[width=0.8\columnwidth]{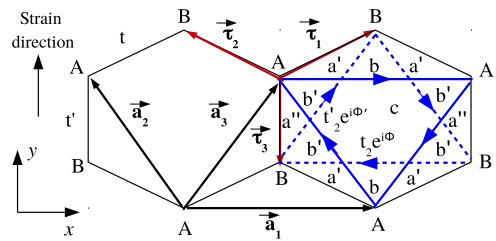}
\caption
{Deformed honeycomb lattice along the armchair $y$ axis.
($\vec{a}_1,\vec{a}_2$) is the lattice basis. 
The hopping parameters to the first (second) neighbors $t$ and $t^{\prime}$ ($t_{2}$ and $t_{2}^{\prime}$) are different due to the deformation. 
Vectors connecting first (second) neighboring atoms are denoted $\vec{\tau}_l$ ($\vec{a}_l$).
The phase pattern for the second-neighbor hopping parameters of the HM is also shown. The arrow indicate the directions along which the hopping integrals $t_2$ and $t_2^{\prime}$ acquire positive phase $e^{i\Phi}$ and $e^{i\Phi^{\prime}}$ respectively. The area of the unit cell is decomposed in regions denoted $a^{\prime}$,$a^{\prime\prime}$, $b$, $b^{\prime}$ and $c$.}
\label{lattice}
\end{figure}

We consider a honeycomb lattice, with two types of atoms ($A$ and $B$), under a uniaxial strain applied along the armchair direction corresponding to the $y$ axis (Fig.\ref{lattice}). In the resulting quinoid lattice, the distance between nearest neighbor atoms, along the strain axis, changes from $a$ to $a^{\prime}=a+\delta a=a(1+\epsilon)$ where 
$\epsilon=\frac{\delta a} a$ is the strain amplitude. For a compressive (tensile) deformation $\epsilon$ is negative 
(positive). It is worth to stress that we only consider the strain component $\epsilon_{yy}=\epsilon$ and neglect, for simplicity, the $\epsilon_{xx}$ term of the strain tensor. This assumption is justified in graphene since the corresponding Poisson ration, relating the strain components $\epsilon_{xx}=-\nu \epsilon_{yy}$, is small ($\nu=0.165$) and decreases with increasing strain amplitude \cite{Poisson16}. Within this assumption, the deformed graphene sheet can be described by the quinoid lattice for which simple analytical expression of the Chern number could be derived, as we shall show in the following.\newline
It should be noted that the quinoid lattice is a good approximation, as far as the strain amplitude is small enough to neglect the strain effect on the bond angles and on the first neighboring distances, along $\vec{\tau}_1$ and  $\vec{\tau}_2$ (Fig.\ref{lattice}).\

The lattice is described by the basis ($\vec{a}_1,\vec{a}_2$) given by: 
\begin{eqnarray}
\vec{a}_1=\sqrt{3}a\vec{e}_x,
\vec{a}_2=-\frac{\sqrt{3}}2a\vec{e}_x+a\left(\frac32 +\epsilon\right)\vec{e}_y,
\end{eqnarray}
The vectors joining the first neighbor atoms are given by:
\begin{eqnarray}
\vec{\tau}_1&=&\frac a 2 \left( \sqrt{3}\vec{e}_x+\vec{e}_y\right),\;
 \vec{\tau}_2=\frac a 2 \left( -\sqrt{3}\vec{e}_x+\vec{e}_y\right),\;\nonumber\\
\vec{\tau}_3&=&-a(1+\epsilon)\vec{e}_y.
\label{tau}
\end{eqnarray}

The second neighboring atoms are connected by the vectors $\pm\vec{a}_1, \pm\vec{a}_2$ 
and $\pm\vec{a}_3=\pm(\vec{a}_1+\vec{a}_2)$.
The hopping integral between first neighboring atoms along $\vec{\tau}_3$ direction is modified by the strain from $t$ to $t^{\prime}=t+\frac{\partial t}{\partial a}\delta a $.\
The hopping terms to the second neighboring atoms $t_{2}$ change also compared to their values in undeformed lattice as: 
\begin{eqnarray}
 t^{\prime}_2=t_2+\frac{\partial t_2}{\partial a}\delta a.
\end{eqnarray}
Assuming the Harrison law, $t^{\prime}$ and $t^{\prime}_2$ could be written as \cite{mark2008,MarkRev}
\begin{eqnarray}
t^{\prime}=t\left(1-2\epsilon\right),\nonumber\\
t^{\prime}_2=t_2\left(1-2\epsilon+\frac{b \epsilon}{2d}\right),
\label{t}
\end{eqnarray}
where $b=\sqrt{3}a$ and for graphene $d=\frac a{3.5}$.

It is worth to note that, the Harrison law is not accurate beyond the linear elastic regime. 
For more accurate values of the hopping amplitudes, Density Functional Theory calculations were proposed \cite{Edouardo15}.

As in the HM, the hopping integrals $t$ and $t^{\prime}$ between first neighboring atoms are real since the paths corresponding to these hopping processes delimit a unit cell with a total zero magnetic flux \cite{Haldane}.
However, the hopping matrix elements $t_2$ and $t^{\prime}_2$ acquire a Peierls phases denoted respectively $\Phi$ and 
$\Phi^{\prime}$. 
Figure \ref{lattice} shows the directions along which the hopping integrals are either $t_2e^{i\Phi}$ or $t^{\prime}_2e^{i\Phi^{\prime}}$.\newline
In general, the complex hopping phases can depend on the strain in different ways, depending on the physical origin of these phases. In the present work, we consider two examples. In the first case, we assume that the phases are proportional to the areas, delimited by triangular contours, through which passes a magnetic flux $\Phi$ or $\Phi^{\prime}$. \newline
The $\Phi$ phase is the same as that of the underformed lattice since the area $S=|\vec{\tau}_1\times\vec{\tau}_2|$ constructed on the vectors $\vec{\tau}_1$ and $\vec{\tau}_2$ in unchanged under the uniaxial strain. However, the phase $\Phi^{\prime}$ may be strain dependent if it is connected to the path delimiting the deformed surface $S^{\prime}$, constructed on the areas $a^{\prime}$, $a^{\prime\prime}$ and $b^{\prime}$, which is deformed under the strain: $\Phi^{\prime}=\frac {2\pi}{\Phi_0}\left(\Phi_{a^{\prime}}+\Phi_{b^{\prime}}+\Phi_{a^{\prime\prime}}\right)$, where $\Phi_{a^{\prime}}$, $\Phi_{b^{\prime}}$, and $\Phi_{a^{\prime\prime}}$ are the fluxes through the $a^{\prime}$, $a^{\prime\prime}$ and $b^{\prime}$ regions and $\Phi_0$ is the flux quantum.
The flux area $S^{\prime}$ could be expressed as a function of the undeformed area $S$ as $S^{\prime}=\frac 12 |\vec{\tau}_1\times\vec{\tau}_3|=\left(1+\epsilon\right)S$, which means that: 
\begin{eqnarray}
\Phi^{\prime}=\left(1+\epsilon\right)\Phi
\label{phip}
\end{eqnarray}
There is also the case where $\Phi^{\prime}=\Phi$ for which the positions of the magnetic fluxes are independent of the deformed lattice as we will discuss in the following. The condition $\Phi^{\prime}\neq \Phi$ is not crucial to tune the topology by the strain as we will show in the following sections.
This point raises the question how to tune the phase $\Phi$ in realistic systems? Is it strain dependent? 
One should expect to realize the strained HM in ultracold atom optical lattices \cite{Esslinger14} which were also used to realize the merging of Dirac cones \cite{Esslingermerge} predicted to occur in graphene under compressive strain \cite{Gilles}. The optical lattice potential parameters could be tuned to mimic the strain effect on the hopping integrals \cite{Esslingermerge}. The staggered fluxes may be controlled, in principle, by the time modulation of the optical lattice. However, the observation of the strain effect on the HM will depend on the parameter range accessible to optical lattice \cite{EsslingerP}.\

The recent proposal of realization of HM in Fe-based honeycomb ferromagnetic insulators \cite{Kee} could be generalized to observe strained HM. Complex hopping integrals, $t_2e^{2i\theta}$, arise in this systems from the $d$ orbitals of Fe ion in the AFe$_2$(PO)$_2$ compounds (A=Ba, K, Cs, La). Applying a strain will change the amplitudes $t_2$ of the $d$ orbital overlappings but not the phases $\theta$ as far as the threefold symmetry is conserved. However, if the bond directions are differently modified by the strain, one should expect a change in the phase factors, which will corresponds to take $\Phi^{\prime}\neq \Phi$, as in our model, but not necessarily following the strain dependence given by Eq.\ref{phip}.\

Chang {\it et al.}\cite{Chang13} reported the observation of QAH effect in thin films of ferromagnetic chromonium doped (Bi,Sb)$_2$Te$_3$. By applying a strain on the film, the hopping integrals are expected to change but not the magnetic fluxes distribution through the unit cell since the magnetic moments will be driven with the atoms by the deformation. This will be ascribed to the case of strain independent phase ( $\Phi^{\prime}=\Phi$).\

The complex phase $\Phi$ could, also, be related to the spin-orbit coupling (SOC) parameter of topological insulators\cite{Guassi}. We shall discuss this issue in Sec.II-C.

The electronic Hamiltonian of the strained lattice can be written, in the basis $\left\{|\Psi^A_{\vec{k}}\rangle,|\Psi^B_{\vec{k}}\rangle\right\}$, associated to the two atoms $A$ and $B$ of the unit cell, as
\begin{eqnarray}
 H(\vec{k})=\left(
\begin{array}{cc}
h_{AA}(\vec{k}) & h^*_{AB}(\vec{k}) \\
 h_{AB}(\vec{k}) & h_{BB}(\vec{k}),
 \label{H2x2}
\end{array}
\right)
\end{eqnarray}
where 
\begin{widetext}
 \begin{eqnarray}
h_{AA}(\vec{k})& = & M+2t_2\left(\cos \Phi \cos\vec{k}.\vec{a}_1+\frac{t_2^{\prime}}{t_2}\cos\Phi^{\prime}(\cos\vec{k}.\vec{a}_2+\cos\vec{k}.\vec{a}_3)\right)
-2t_2\left(\sin \Phi \sin\vec{k}.\vec{a}_1+\frac{t_2^{\prime}}{t_2}\sin\Phi^{\prime}(\sin\vec{k}.\vec{a}_2-\sin\vec{k}.\vec{a}_3)\right)\nonumber \\ 
h_{BB}(\vec{k})&=&-M+2t_2\left(\cos \Phi \cos\vec{k}.\vec{a}_1+\frac{t_2^{\prime}}{t_2}\cos\Phi^{\prime}(\cos\vec{k}.\vec{a}_2+\cos\vec{k}.\vec{a}_3)\right) +2t_2\left(\sin \Phi \sin\vec{k}.\vec{a}_1+\frac{t_2^{\prime}}{t_2}\sin\Phi^{\prime}(\sin\vec{k}.\vec{a}_2-\sin\vec{k}.\vec{a}_3)\right) \nonumber\\
 h_{AB}(\vec{k})&=&t^{\prime} e^{i\vec{k}.\vec{\tau}_3}+t( e^{i\vec{k}.\vec{\tau}_1}+ e^{i\vec{k}.\vec{\tau}_2})
 \label{hbb}
\end{eqnarray}
\end{widetext}
$H(\vec{k})$ can be expressed, using the $2\times 2$ Pauli matrices $\vec{\sigma}=\sigma_1\vec{e}_x+\sigma_1\vec{e}_y+\sigma_3\vec{e}_z$ and the identity matrix $\sigma^0={1\!\!1}$, as:
\begin{equation}
H(\vec{k})=h_0(\vec{k})\sigma^0+\sum_{i=1}^3 h_{i}(\vec{k})\sigma^{i},
\label{Hgeneral}
\end{equation}
with
\begin{eqnarray}
h_{0}(\vec{k})&=&2t_2\left(\cos \Phi \cos\vec{k}.\vec{a}_1+\frac{t_2^{\prime}}{t_2}\cos\Phi^{\prime}(\cos\vec{k}.\vec{a}_2+\cos\vec{k}.\vec{a}_3)\right)  \nonumber\\
 h_{x}(\vec{k})&=&t^{\prime} \cos (\vec{k}.\vec{\tau}_3)+t(\cos \vec{k}.\vec{\tau}_1+ \cos\vec{k}.\vec{\tau}_2) \nonumber \\
 h_{y}(\vec{k})&=&t^{\prime} \sin (\vec{k}.\vec{\tau}_3)+t(\sin \vec{k}.\vec{\tau}_1+ \sin\vec{k}.\vec{\tau}_2)  \nonumber \\
h_{z}(\vec{k})&=& M- 2t_2\left(\sin \Phi \sin\vec{k}.\vec{a}_1+\frac{t_2^{\prime}}{t_2}\sin\Phi^{\prime}(\sin\vec{k}.\vec{a}_2-\sin\vec{k}.\vec{a}_3)\right)\nonumber \\
\label{h0}
\end{eqnarray}
The Hamiltonian given by Eq.\ref{Hgeneral} is invariant under time reversal transformation if
\begin{widetext}
\begin{eqnarray}
H^*(-\vec{k})&=&H(\vec{k})+2 \left[2t_2\left(\sin \Phi \sin\vec{k}.\vec{a}_1+\frac{t_2^{\prime}}{t_2}\sin\Phi^{\prime}(\sin\vec{k}.\vec{a}_2-\sin\vec{k}.\vec{a}_3\right)\right]\sigma_3=H(\vec{k}),
\label{TRS}
\end{eqnarray}
\end{widetext}

which yields to the condition: $\sin\Phi=0$ \textit{and} $\sin\Phi^{\prime}=0$.
We then expect, that under strain, the trivial insulating state for $\Phi=\pi$ of the undeformed lattice, could be tuned to a topological phase if $\Phi^{\prime}\neq\Phi$ (Eq.\ref{phip}). The topology of the HM under strain is, then, not only dependent on the local magnetic flux, but also on the strain amplitude. The question is whether the strain competes with the topology. The answer will be given in the next sections.\\

The eigenvalues of the Hamiltonian given by Eq.\ref{Hgeneral} are:
\begin{equation}
\epsilon_{\lambda}(\vec{k})=h_{0}(\vec{k})+\lambda |h(\vec{k})|,
\end{equation}

where $\lambda=\pm$ is the band index.\
For $M=0$ and $\Phi=0$, one recovers the band structure of graphene under a uniaxial strain showing two bands touching at the Dirac points $D$ and $D^{\prime}$ given by $\vec{k}_{D,D^{\prime}}=\left(k^{\xi}_{Dx},0\right)$, where the component $k^{\xi}_{Dx}$ at the valley $\xi=\pm$ is given by \cite{mark2008}
\begin{eqnarray}
 k^{\xi}_{Dx}=\xi \frac2{\sqrt{3}a}\arccos\left(-\frac{t^{\prime}}{2t}\right).
 \label{k_D}
\end{eqnarray}
Under the strain, the Dirac cones leave the high symmetry points $K$ and $K^{\prime}$ and move towards each other, under a compressive strain ($\epsilon<0$) and can, eventually, merge for $\epsilon=-0.5$ \cite{mark2008}.\

To study the topological character of the HM under strain, one needs to determine the corresponding Chern number whose analytical expression could be derived taking the low energy form of the Hamiltonian of Eq.\ref{Hgeneral}, the so-called continuum limit.

\subsection{Chern number: continuum limit}

The Hamiltonian given by Eq.\ref{Hgeneral} could be expanded around the Dirac points as:
\begin{eqnarray}
H_{\xi}(\vec{q})=\left(
\begin{array}{cc}
m_{\xi}+\xi \hbar w_{0x}q_x &\xi\hbar(w_{x} q_x-i\xi w_{y} q_y)\\
 \xi\hbar (w_{x} q_x+i\xi w_{y} q_y)&-(m_{\xi}-\xi \hbar w_{0x}q_x)
\end{array}
\right),\nonumber\\
\label{HD}
\end{eqnarray} 

where $w_{x}$ and $w_{y}$ are the anisotropic Fermi velocities and $w_{0x}$ is the tilt parameter, given by:

\begin{eqnarray}
w_x&=&\frac 32 \frac{at}{\hbar} \left(1+\frac 23 \epsilon\right),\,w_y=\frac 32 \frac{at}{\hbar} \left(1-\frac 43 \epsilon\right)\nonumber\\
w_{0x}&=& -\frac {2\sqrt{3}a}{\hbar}\left(t_2 \cos\Phi \sin 2\theta+t^{\prime}_2  \cos\Phi^{\prime} \sin\theta\right),
\label{tilt}
\end{eqnarray}
where $\theta$ is defined as:
\begin{eqnarray}
 \theta=\arccos\left(-\frac{t^{\prime}}{2t}\right)
 \label{theta}
\end{eqnarray}

The mass term $m_{\xi}$ is :
\begin{equation}
 m_{\xi}=M+\xi 2t_2\left(\frac{2t^{\prime}_2}{t_2}\sin\Phi^{\prime} \sin\theta -\sin\Phi \sin 2\theta\right)
 \label{mass}
\end{equation}
The tilt term is obtained by expanding$h_0(\vec{k})$, to the first order, around $\vec{k^{\xi}_D}=(k^{\xi}_{Dx},k^{\xi}_{Dy}=0)$, where $k^{\xi}_{Dx}$ is given by Eq.\ref{k_D}. The mass term corresponds to the zeroth order expansion of $h_z(\vec{k})$. We have neglected the first order term, which is valley independent, and then renormalizes equally the Fermi velocity along the $q_x$ axis in both valleys.

The dispersion relation reduces to:
\begin{equation}
\epsilon_{\lambda}^{\xi}(\vec{q})=\xi\hbar w_{0x}q_x+\lambda\hbar \sqrt{w_{x}^2 q_x^2+ w_{y}^2q_y^2+m_{\xi}^2},
\label{disper}
\end{equation}
which describes massive Dirac fermions moving with anisotropic velocities along the $x$ and $y$ directions.\\

The topological character of a phase is determined by the first Chern number given by $\mathcal C=\mathcal C_{\xi}+\mathcal C_{-\xi}$, where $\mathcal C_{\xi}$ is the Chern number calculated at the valley $\xi$.\newline
To derive an analytical expression of the Chern number, we neglect the tilt term $w_{0x}$ since it does not change the topological character of the system as it contributes, in the Hamiltonian (Eq.\ref{HD}), with the identity matrix in each valley. The Hamiltonian around the Dirac points (Eq.\ref{HD}), could then be written as:
\begin{equation}
H_{\xi}(\vec{q})=\vec{h}_{\xi}(\vec{q}).\vec{\sigma},
\end{equation}
with $\vec{h}_{\xi}(\vec{q})=\left(\xi w_xq_x,w_yq_y,m_{\xi}\right)\equiv|\vec{h}_{\xi}(\vec{q})|\vec{n}_{\xi}(\vec{q}) $.

The corresponding Chern number reads as:
\begin{equation}
\mathcal C_{\xi} = \frac {1}{2\pi}\int \Omega_{\xi} (\vec{q}) d^2\vec{q},
\end{equation}
where $\Omega_{\xi}$ is the component of the Berry curvature along the unitary vector $\vec{n}_{\xi}(\vec{q})$: $\Omega_{\xi}=\frac 12 \left[ \partial_{q_x}\vec{n}_{\xi}(\vec{q})\times\partial_{q_y}\vec{n}_{\xi}(\vec{q})\right].\vec{n}_{\xi}(\vec{q}) $ \cite{BookBerry}.\

Straightforward calculations give:
\begin{equation}
\mathcal C = \frac {1}{2}\left[ \sign(m_{+})-\sign(m_{-})\right],
\label{chern}
\end{equation}
This expression is reminiscent of that obtained in the HM in the undeformed lattice. However the mass terms are, now, strain dependent (Eq.\ref{mass}). \
In the following, we discuss the corresponding phase diagram.\\

\subsection{Haldane model under strain: Phase diagram}

\begin{figure}[hpbt] 
\begin{center}
\includegraphics[width=0.7\columnwidth]{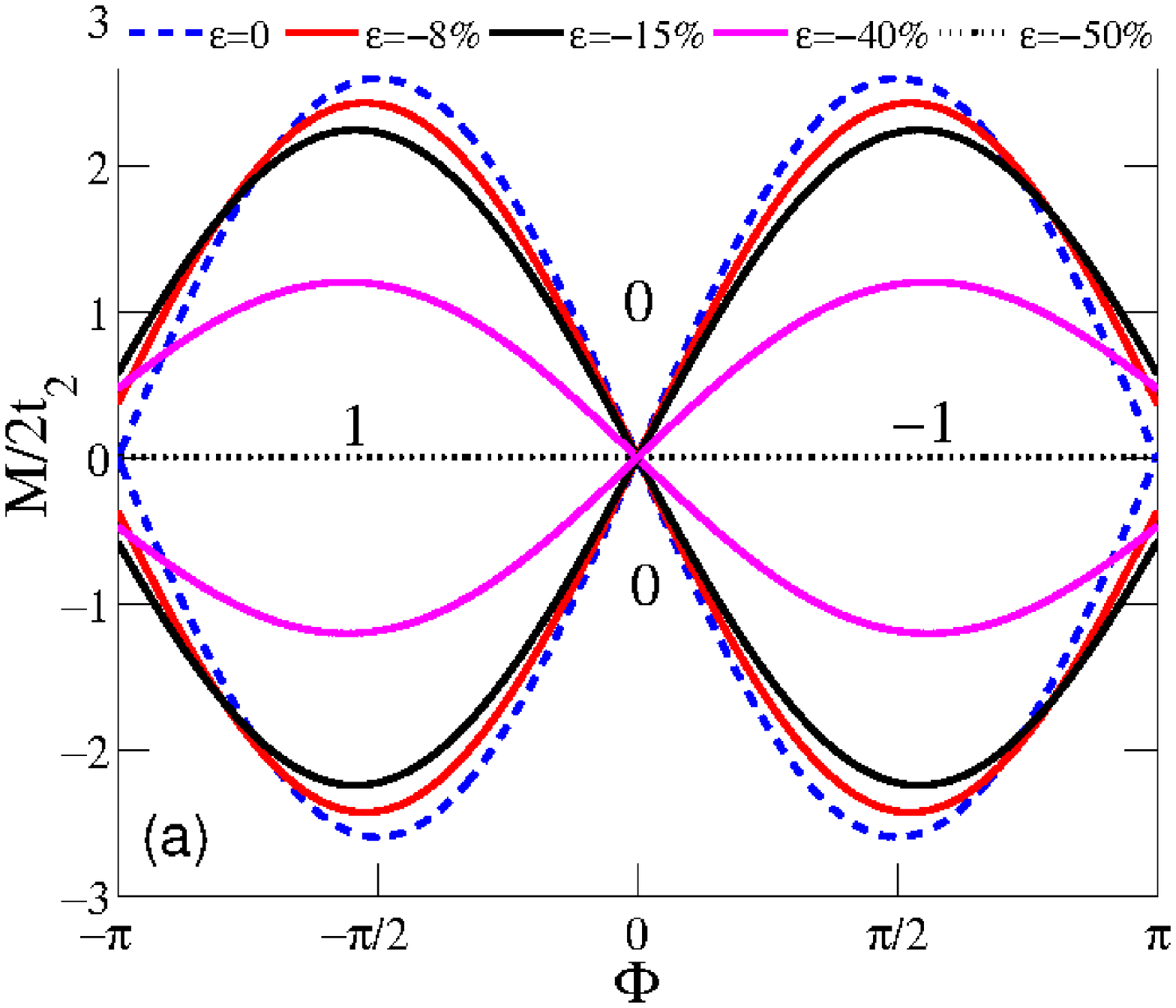}
\includegraphics[width=0.7\columnwidth]{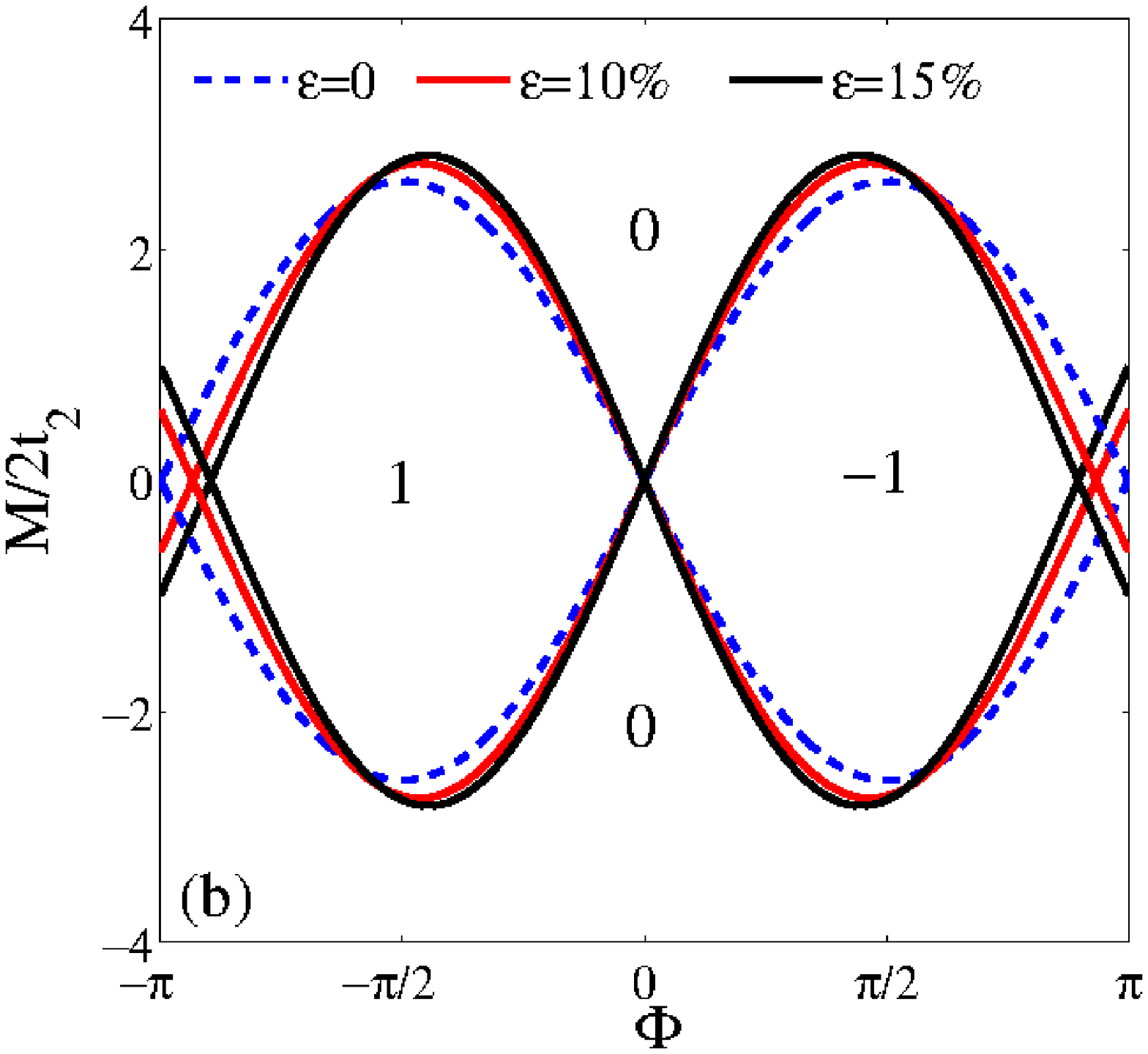}
\end{center}
\caption{Phase diagram of the HM for different strain amplitudes. The case of compressive (tensile) strain is shown in the upper (lower) panel. Calculations are done for $t_2=0.1 t$ and for $\Phi^{\prime}=(1+\epsilon)\Phi$ (Eq.\ref{phip}).}
\label{diagram}
\end{figure}

We first discuss the case where $\Phi^{\prime}$ is strain dependent (Eq.\ref{phip}). In figure \ref{diagram}, we represent the phase diagram of the HM under uniaxial strain as a function of $\frac M{2t_2}$ and $\Phi$ for different strain values. The calculations are done for $t_2=0.1 t$.
The phase boundaries between the trivial ($\mathcal C =0$) and the topological ($\mathcal C =\pm 1$) phases corresponds to the case where one Dirac points is gaped ($m_{\xi}\neq 0$) and the other is not ($m_{-\xi}=0$), which yields to the condition:
\begin{eqnarray}
 \frac {M}{2t_2} =\pm |\sin\Phi\sin 2\theta-\frac{2t_2^{\prime}}{t_2}\sin\Phi^{\prime}\sin \theta|,
 \label{transit-cond}
\end{eqnarray}
where $\theta$ and $\Phi^{\prime}$ are strain dependent (Eqs.\ref{phip} and \ref{theta}).\\

Figure \ref{diagram} shows that, under strain, the $2\pi$ periodicity of the Haldane phase diagram is not conserved in the case where $\Phi^{\prime}=(1+\epsilon)$ (Eq.\ref{phip}), which results in a nonvanishing Chern number for $\Phi=\pi$ and $M = 0$ (Fig.\ref{diagram}). This is due to the strain dependence periodicity of $\sin\Phi^{\prime}$ which is $2\pi/(1+\epsilon)\sim 2\pi(1−\epsilon)$. As shown in Fig.\ref{diagram}, the pseudo-periodicity of the transition line (increases) for a tensile (compressive) deformation compared to the undeformed case.
The strain dependent pseudoperiodicity of the HM phase diagram could be probed in optical lattices \cite{Esslinger14}.\

At the critical value of $\epsilon=-0.5$, the system turns to a trivial insulator since the Dirac cones merge for this strain amplitude and the electrons loose their Dirac character \cite{mark2008}.\

\begin{figure}[hpbt] 
\begin{center}
\includegraphics[width=0.8\columnwidth]{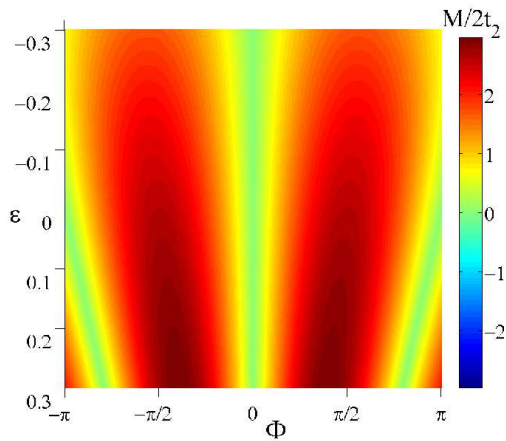}
\end{center}
\caption{Evolution of the topological phases of the HM under strain. The point $(M=0,\Phi=0)$, at which the transition between two topological phases take place in the undeformed lattice, is found to be shifted by the strain to $(M=0,\Phi\neq0)$. Calculations are done for $t_2=0.1 t$ and for $\Phi^{\prime}=(1+\epsilon)\Phi$ (Eq.\ref{phip}.}
\label{map}
\end{figure}

According to figure \ref{diagram}, the strain could drive a topological phase, of the undeformed lattice, into a trivial one. In particular, at a tensile strain of $\epsilon=0.15$, the topological phase with $\mathcal C=-1$ (for $M=t_2$ and $\Phi=\frac {31}{36}\pi$) switches to a trivial phase ($\mathcal C=0$). On the other hand, a topological state with $\mathcal C=-1$ ($M=0$ and $\Phi=\frac {14}{15}\pi$) could turn to an other topological phase with an opposite Chern number by applying a tensile deformation of the order of $\epsilon=0.15$. 
These results are summarized in figure \ref{map} where we depicted the phase boundaries as a function of $\Phi$ and the strain amplitude $\epsilon$. The color map indicates the value of the sublattice potential $M$ at which a phase transition takes place. This figure shows that the phase boundaries are strain dependent and that a compressive strain compete with the non-trivial topological character of the system. However, a tensile deformation furthers the formation of topological states and the transitions between phases with opposite Chern numbers.
These features may lead to a strain tuned topology with switchable edge currents, which could be of a great interest for quantum computing \cite{TIqucomp,Liu,Akiho}. Recently, a single-photon emitter, a key component for computing devices, was realized based on strain engineering of a topological 2D materials (WSe$_2$) \cite{Rosenberger}.\

\begin{figure}[hpbt] 
\begin{center}
\includegraphics[width=0.7\columnwidth]{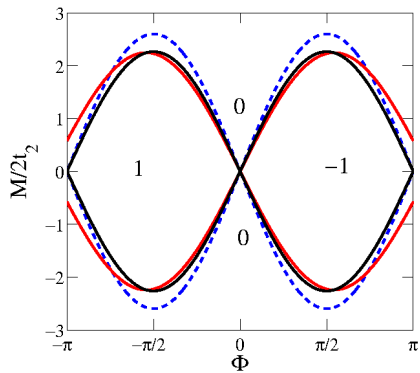}
\end{center}
\caption{Phase diagram of the HM for a compressive strain of $\epsilon=-0.15$ and $t_2=0.1t$. 
The dashed line corresponds to the undeformed lattice. The black (red) line is the phase boundary in the case where $\Phi^{\prime}=\Phi$ ($\Phi^{\prime}=(1+\epsilon)\Phi$, Eq.\ref{phip}).}
\label{diagram2}
\end{figure}

Actually, according to Eq.\ref{transit-cond}, the strain dependence of the phase boundaries in figure \ref{diagram} is not only due to the presence of different phase $\Phi$ and $\Phi^{\prime}$ (Eq.\ref{phip}). The Haldane phase diagram will be affected by the uniform uniaxial strain even if $\Phi=\Phi^{\prime}$, since the line boundaries will depend on the strain amplitude through the ratio $t_2^{\prime}/t_2$ (Eq.\ref{transit-cond}).
This feature is shown in figure \ref{diagram2} where we depicted the Haldane phase diagram under a uniaxial strain in the cases where $\Phi=\Phi^{\prime}$ and $\Phi\neq\Phi^{\prime}$. This figure shows that, taking $\Phi=\Phi^{\prime}$ restores the $2\pi$ periodicity of the Haldane phase diagram. However, the topology is still affected by the strain as in the case where $\Phi\neq\Phi^{\prime}$ (Eq.\ref{phip}).

\begin{figure}[hpbt] 
\begin{center}
$\begin{array}{cc}
\includegraphics[width=0.5\columnwidth]{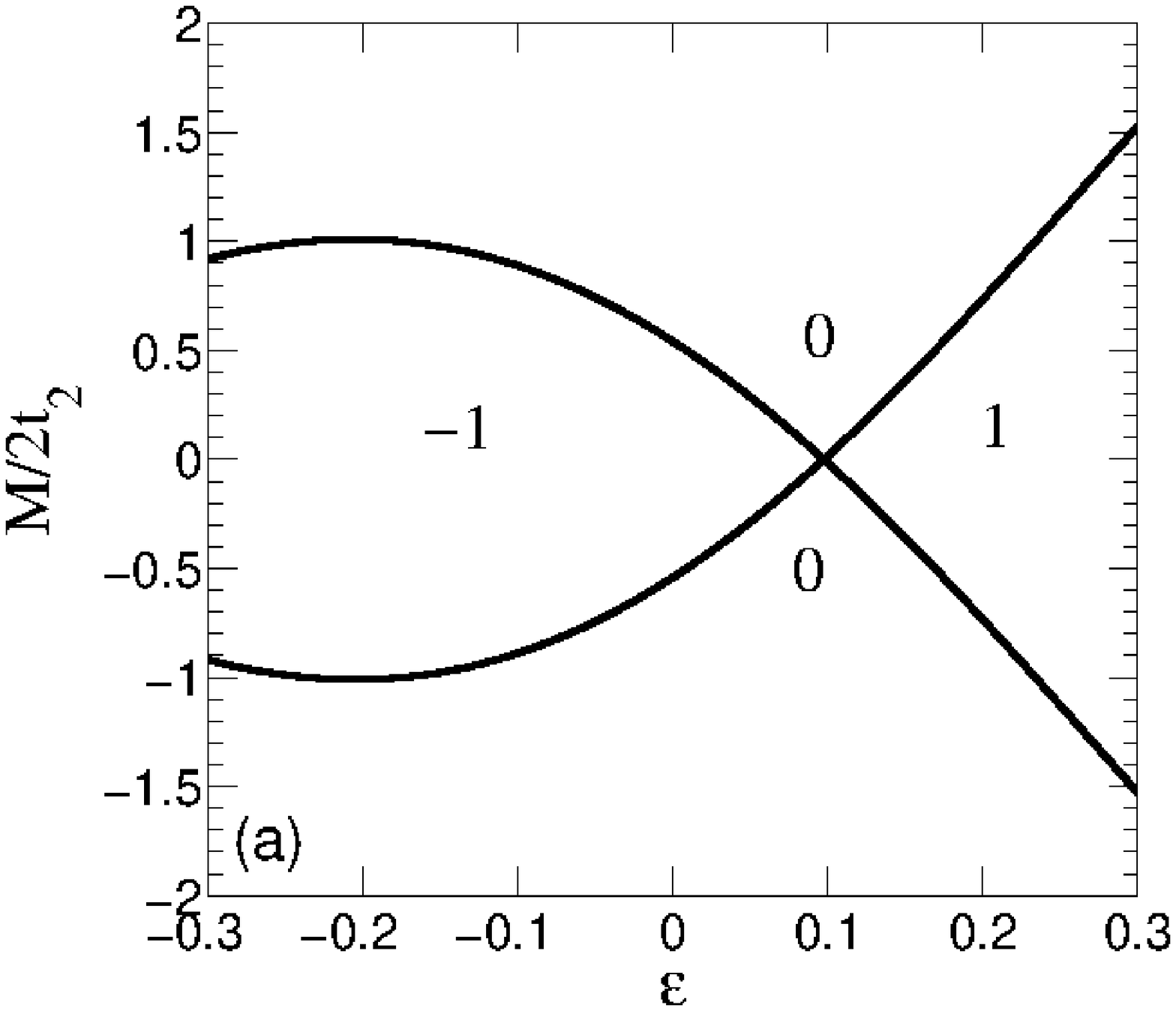}&
\includegraphics[width=0.5\columnwidth]{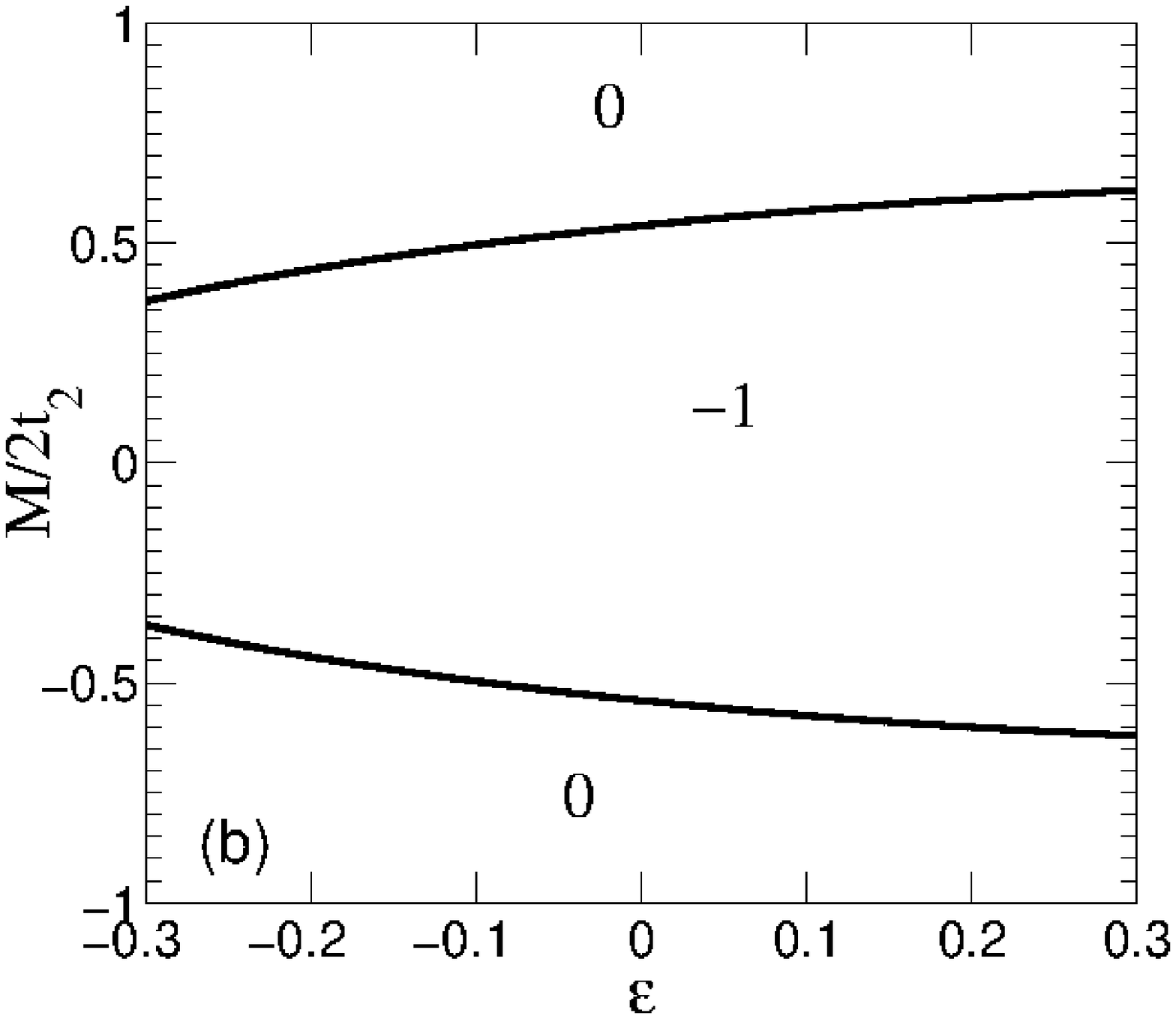}
\end{array}$
\end{center}
\caption{Phase diagram of the HM as a function of the strain amplitude for $\Phi=\frac{14}{15}\pi$ and $t_2=0.1t$. Calculations are done for (a) $\Phi^{\prime}=(1+\epsilon)\Phi$ (Eq.\ref{phip}) and (b) $\Phi=\Phi^{\prime}$.}
\label{M_eps}
\end{figure}

The transition boundary line for $\Phi=\frac{14}{15}\pi$ of figure \ref{diagram} is represented in figure \ref{M_eps} as function of the strain amplitude.
The figure shows, that at a given non vanishing mass value $M$, the system could undergo transitions between phases with different Chern numbers by tuning the strain. Moreover, the strain may change the sign of the Chern number of a given topological phase. As a consequence, the corresponding edge currents direction of propagation is expected to be switchable by the strain, which may pave the way to the strain engineering of the edge currents.\\

In figure \ref{diagram}, the line boundaries correspond to the regime where the topological gap opens in one valley and closes in the other. This feature can be brought out in the band structure calculated within the tight binding approach for a zigzag nanoribbon as we shall discuss in section III (Fig.\ref{TB}).

\begin{figure}[hpbt] 
\begin{center}
\includegraphics[width=0.5\columnwidth]{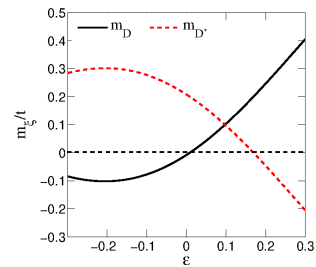}
\end{center}
\caption{Strain dependence of the topological gaps of the Haldane model (Eq.\ref{mass}) for $t_2=0.1t$, $M=t_2$ and $\phi=\frac{14}{15}\pi$ at the Dirac points $D$ and $D^{\prime}$. Calculations are done for $\Phi^{\prime}\neq\Phi$ (Eq.\ref{phip}.}
\label{gap}
\end{figure}

The strain dependence of this gap is depicted in figure \ref{gap} which shows that a uniform uniaxial strain could tune the topological gap. This behavior is different from that found in the case of HM under a nonuniform strain where the gap is found to be weakly modified by the strain \cite{Ghaemi,Castro17}

\begin{figure}[hpbt] 
\begin{center}
$\begin{array}{cc}
\includegraphics[width=0.5\columnwidth]{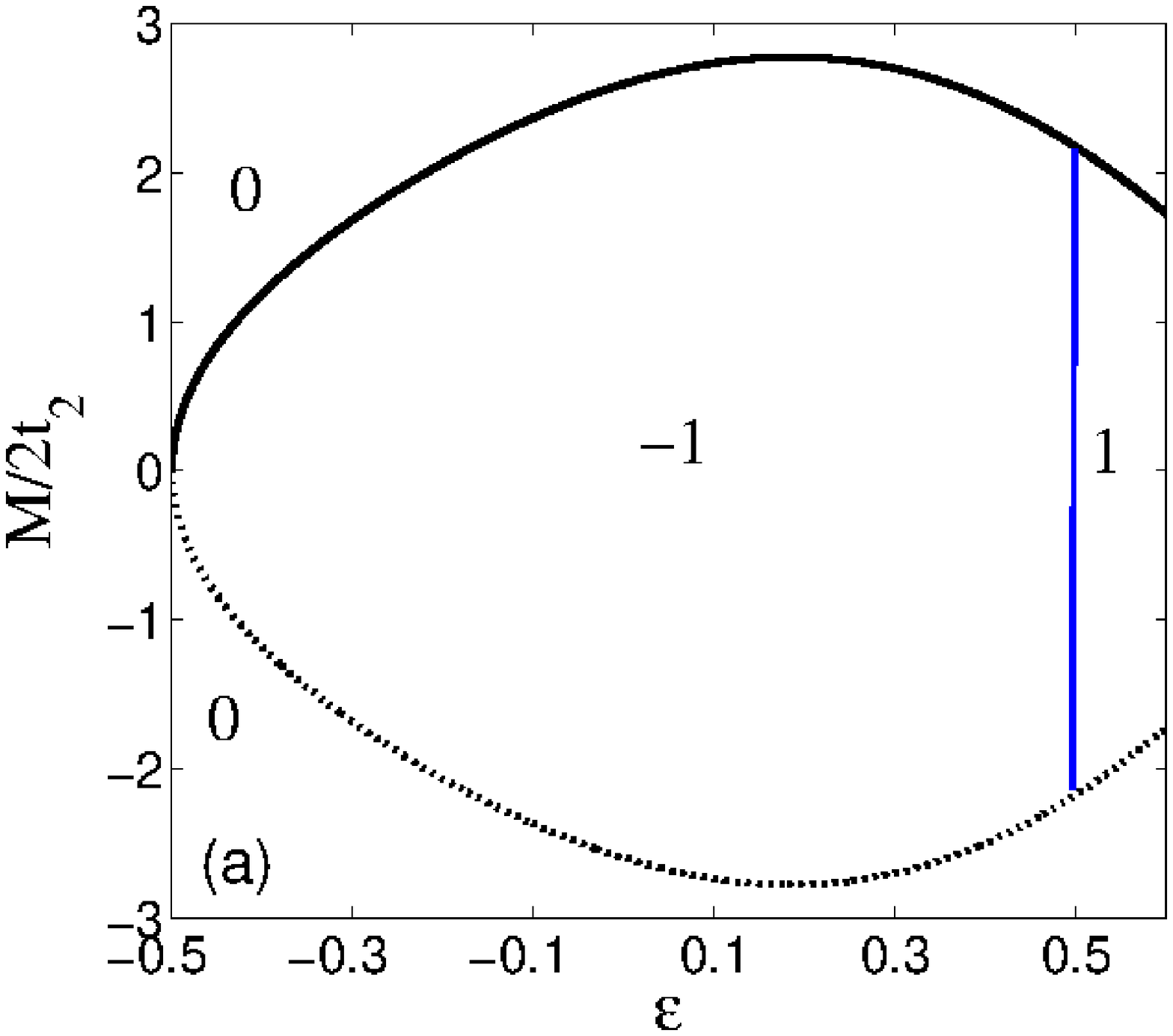}&
\includegraphics[width=0.5\columnwidth]{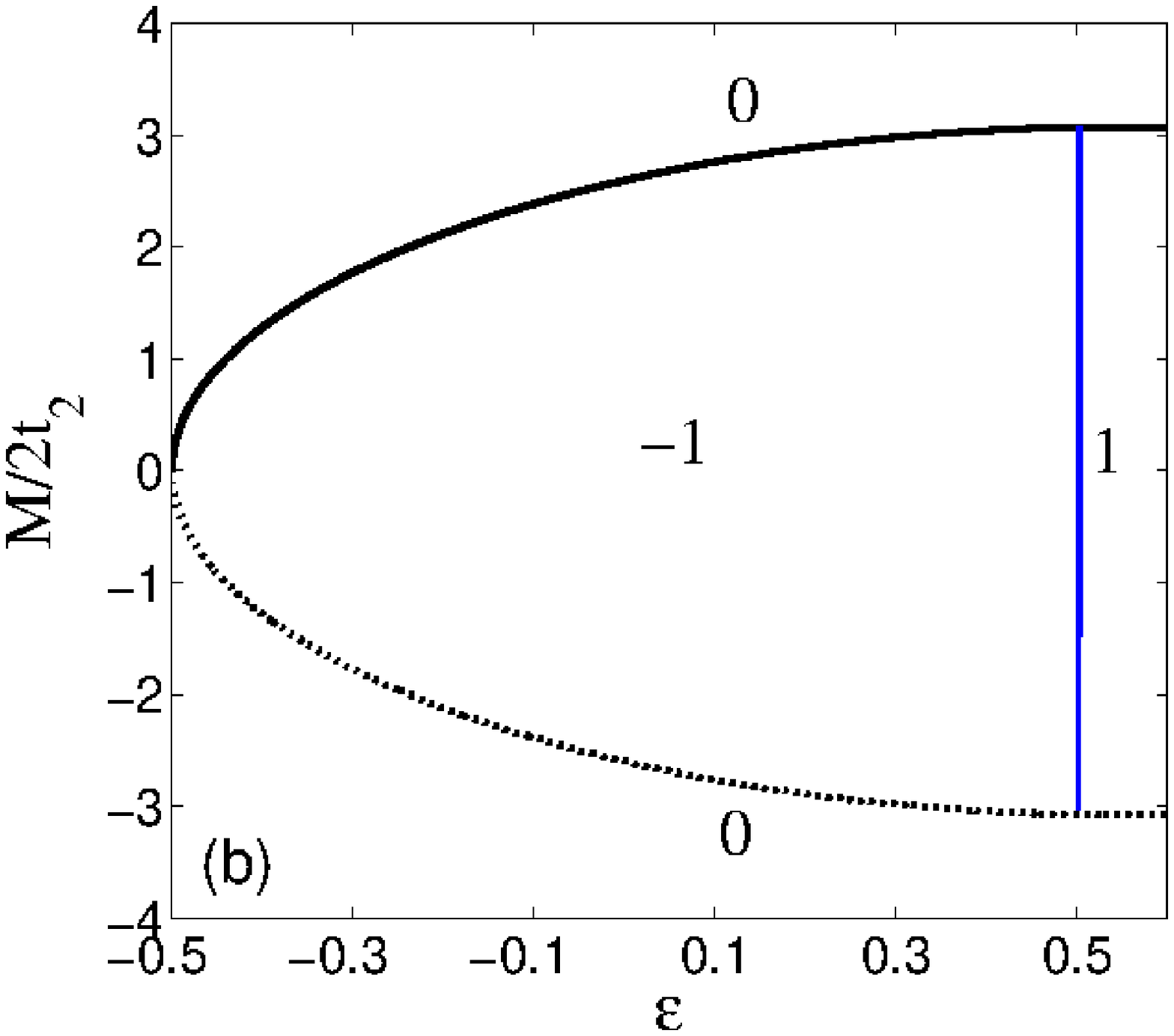}
\end{array}$

\end{center} 
\caption{HM phase diagram as a function of the strain amplitude up to the large strain regimes. The system becomes trivial at the critical value $\epsilon=-0.5$ at which the Dirac cones merge. A boundary line between two topological phases with opposite Chern numbers is found at $\epsilon=0.5$ for which the hopping term, to the first neighboring atoms, along the strain direction vanishes. Calculations are done for $\Phi=\frac{\pi}2$, $t_2=0.1t$, and for (a) $\Phi^{\prime}=(1+\epsilon)\Phi$ (Eq.\ref{phip}) and (b) $\Phi=\Phi^{\prime}$.}
\label{merging}
\end{figure}

It is interesting to address the behavior of the topological phases when the Dirac cones merge at the critical strain value of $\epsilon=-0.5$ (Eq.\ref{k_D}). We plot, in figure \ref{merging}, the evolution, as a function of strain, of the  phase boundary of the topological state obtained at $\Phi=\frac{\pi}2$. This figure shows that at the merging point, the system turns to a trivial state as it is expected \cite{Gilles}.

On the other hand, at the tensile strain amplitude $\epsilon=0.5$, transitions between topological phases with opposite Chern numbers take place at finite $M$, which results in a line boundary separating the phases $C=1$ and $C=-1$. In the undeformed HM, such transition occurs only at the point $M=0$ and $\Phi=0$. This feature is reminiscent of the result found in the case of the Chern insulator on a square lattice \cite{square} and in disordered semi-Dirac material \cite{Moessner}. 
The presence of this transition could be understood from the expression of the Hamiltonian given by Eq.\ref{Hgeneral} at the critical value $\epsilon=0.5$ for which the hopping parameter $t^{\prime}$ vanishes (Eq.\ref{t}).
Disregarding the diagonal term $h_0(\vec{k})$, which does not affect the topology of the HM, the Hamiltonian reduces to $h_z(\vec{k})$.
The Dirac points, defined by $h_z(\vec{k}_D)=0$, satisfy:
\begin{eqnarray}
\vec{k}_D=\left(k_{Dx}=\xi \frac{\pi}{\sqrt{3}a},k_{Dy}=\frac{1}{2a}\arccos\left(-\xi\frac M {4t^{\prime}_2\sin\Phi^{\prime}}\right)\right),\nonumber \\
\end{eqnarray}
which gives rise to the condition: 
\begin{eqnarray}
 |M|\le |4t^{\prime}_2\sin\Phi^{\prime}|,
 \label{mphi}
\end{eqnarray}
For these $M$ values, the gap closes at one of the Dirac points and the system undergoes a transition between tow topological phases ($C=-1$ and $C=1$), which results in a line boundary separating the two phases as shown in figure \ref{merging}. The result holds for $\Phi^{\prime}=\Phi$ and $\Phi^{\prime}\neq\Phi$.
Such phase boundary could be observed in cold atoms trapped in optical lattices if the lattice parameters could be tuned to reach the extreme strain amplitude regime \cite{Esslinger14,Esslingermerge}.
It should be stressed that the strain value $|\epsilon|=0.5$ is large and the Harrison law is no more accurate at this regime, as mentioned in section II-A.

The generalization of the present work to spinfull systems, may provide insights into the behavior of the edge states of topological insulators subjected to a uniaxial strain \cite{Kane11}. Actually, the strain dependence of the phase $\Phi^{\prime}$, is reminiscent of the strain dressed intrinsic SOC of a graphene nanoribbon \cite{Guassi}, where the low SOC regime corresponds to the QAH state.
Guassi {\it et al.}\cite{Guassi} showed that the strain-induced pseudomagnetic field couples to the spin degrees of freedom in deformed graphene, which results in a strain dressed SOC parameter. 
The dependence of the SOC term of the Hamiltonian (Eq.1) of Ref.\onlinecite{Guassi} on the cross product of $\vec{\tau}_i$ vectors is reminiscent of the expressions of the phases $\Phi$ and $\phi^{\prime}$, we assumed in the present work, and which are related to the triangular areas constructed on the $\vec{\tau}_i$ vectors.\

It is worth to note that the phase diagram of Fig.\ref{diagram} is derived within the continuum approximation, which is not accurate beyond the low energy limit\cite{Castro17}. We then present, in the following, a tight-binding (TB) approach to discuss the role of a uniaxial strain on the edge states of a nanoribbon described by the HM.

\section{Haldane model under strain: a tight-binding approach}

\begin{figure}[hpbt] 
\begin{center}
\includegraphics[width=0.6\columnwidth]{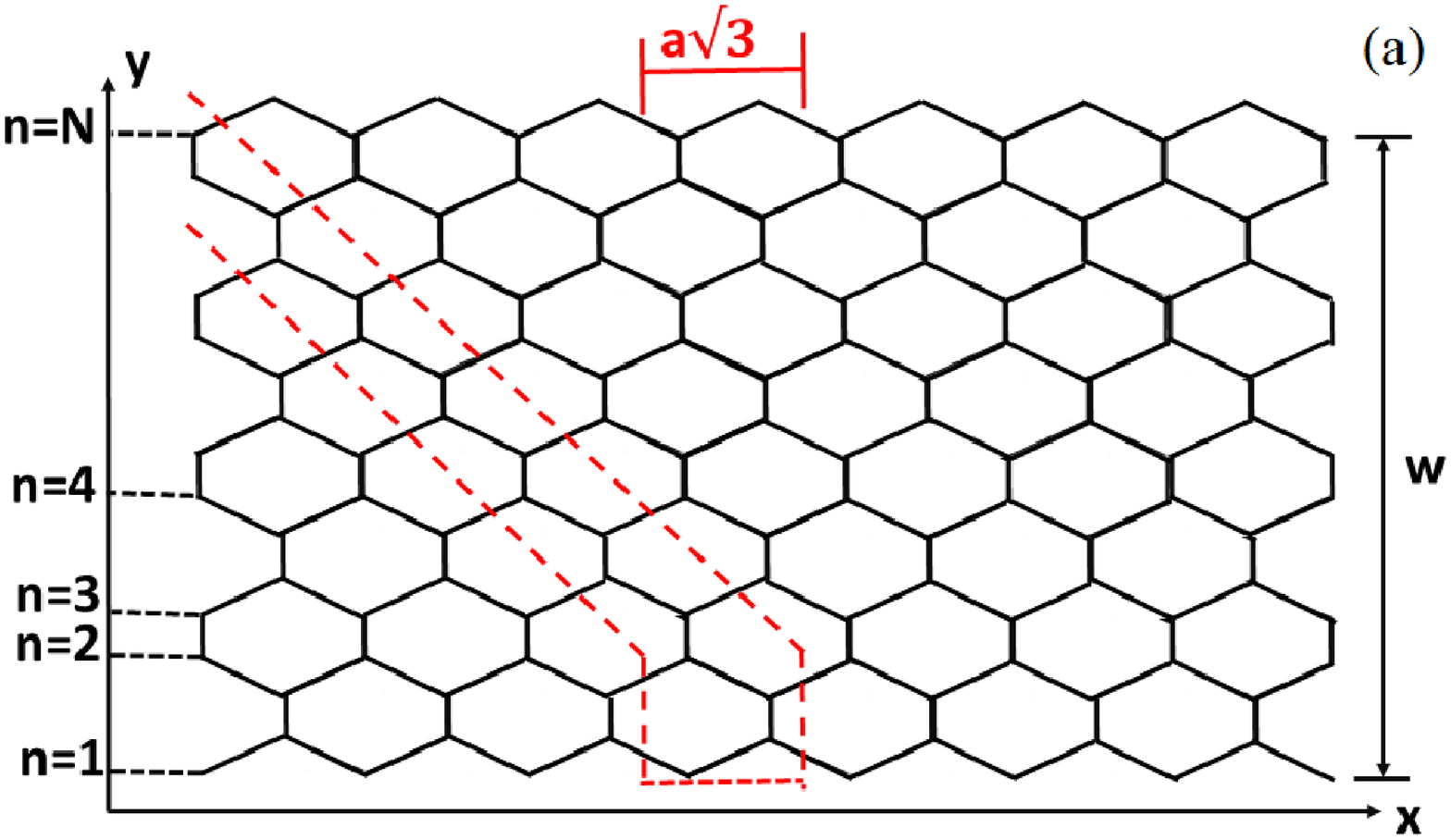}
$\begin{array}{cc}
\includegraphics[width=0.5\columnwidth]{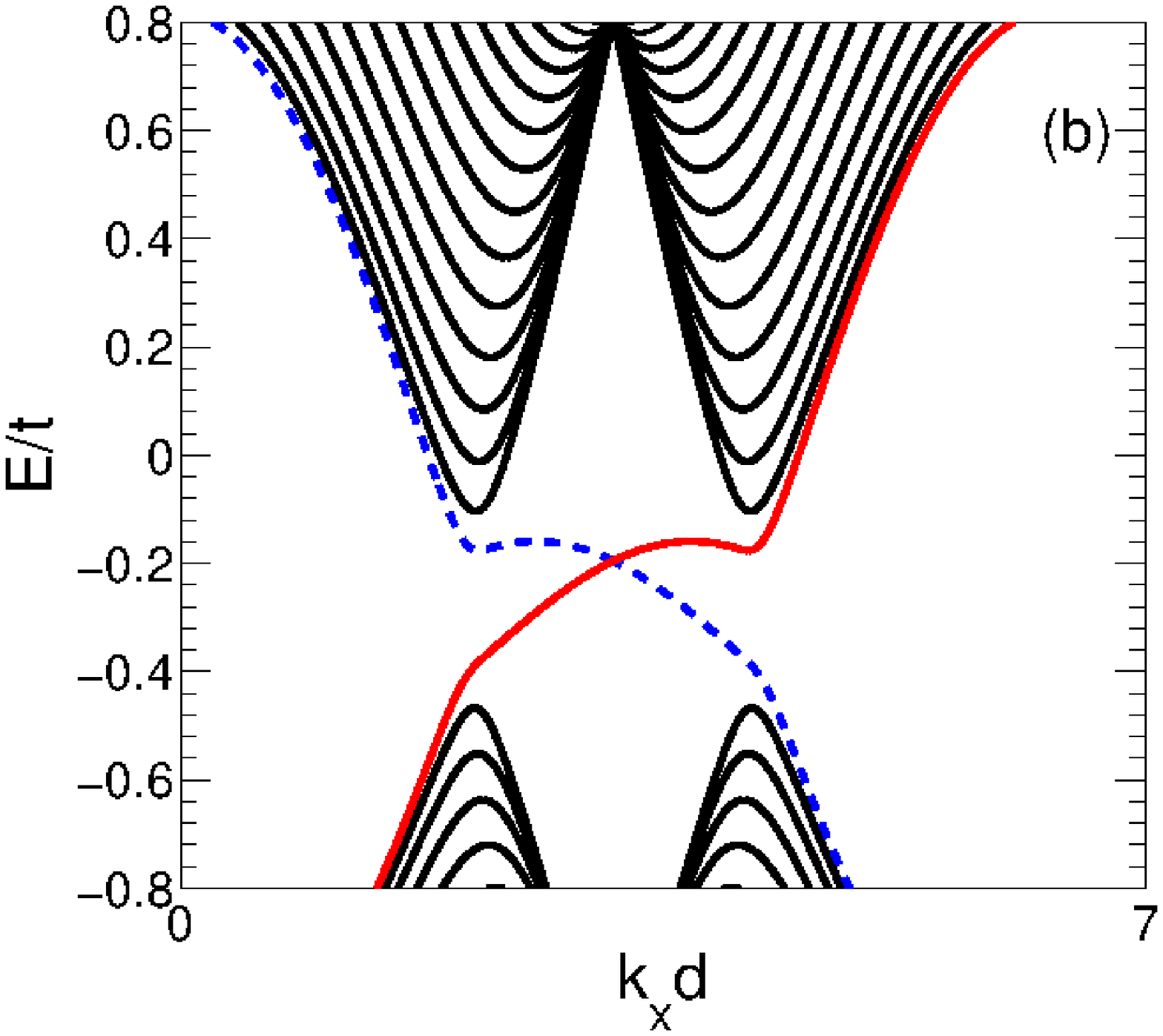}&
\includegraphics[width=0.5\columnwidth]{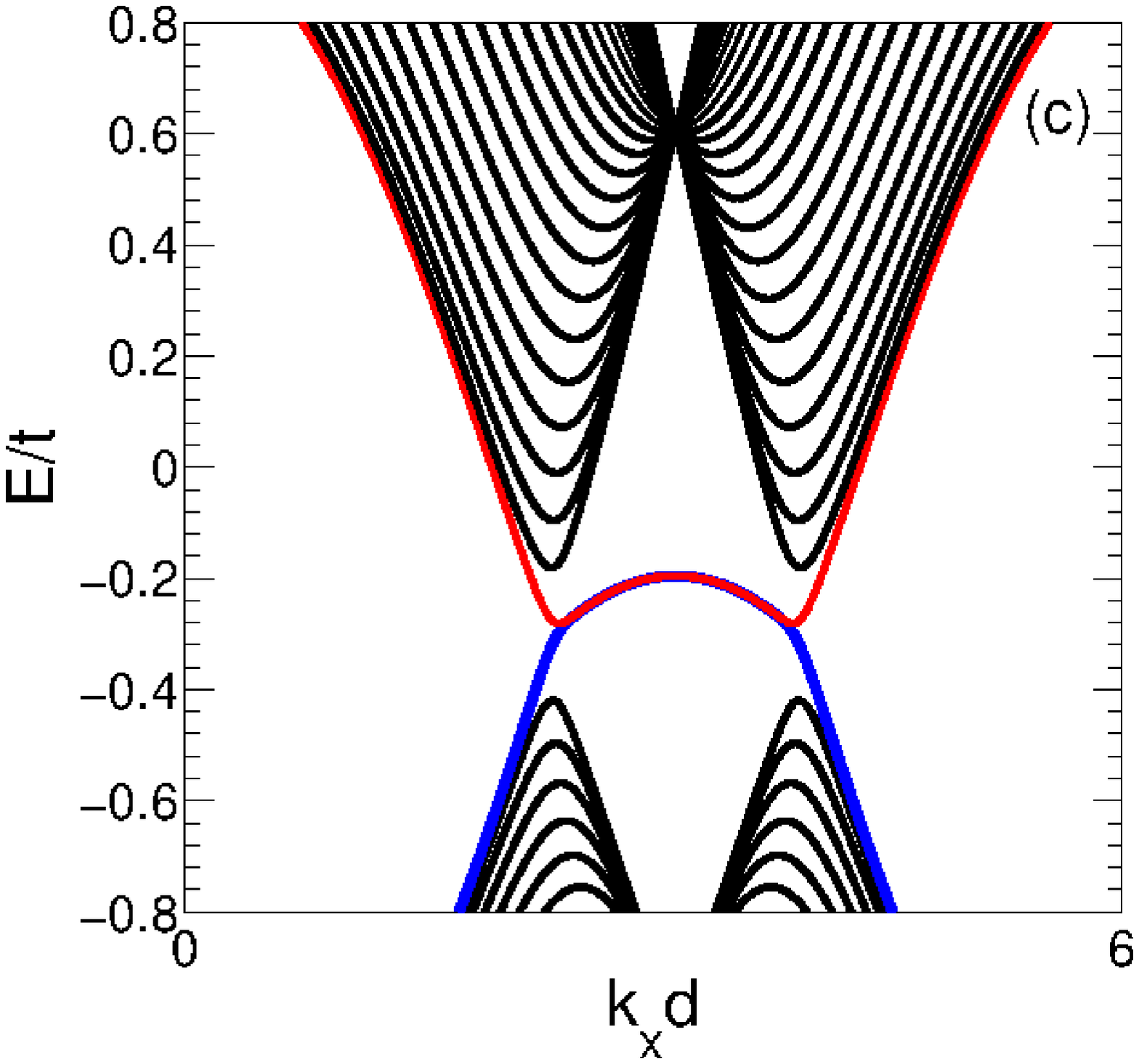}
\end{array}$
$\begin{array}{cc}
\includegraphics[width=0.5\columnwidth]{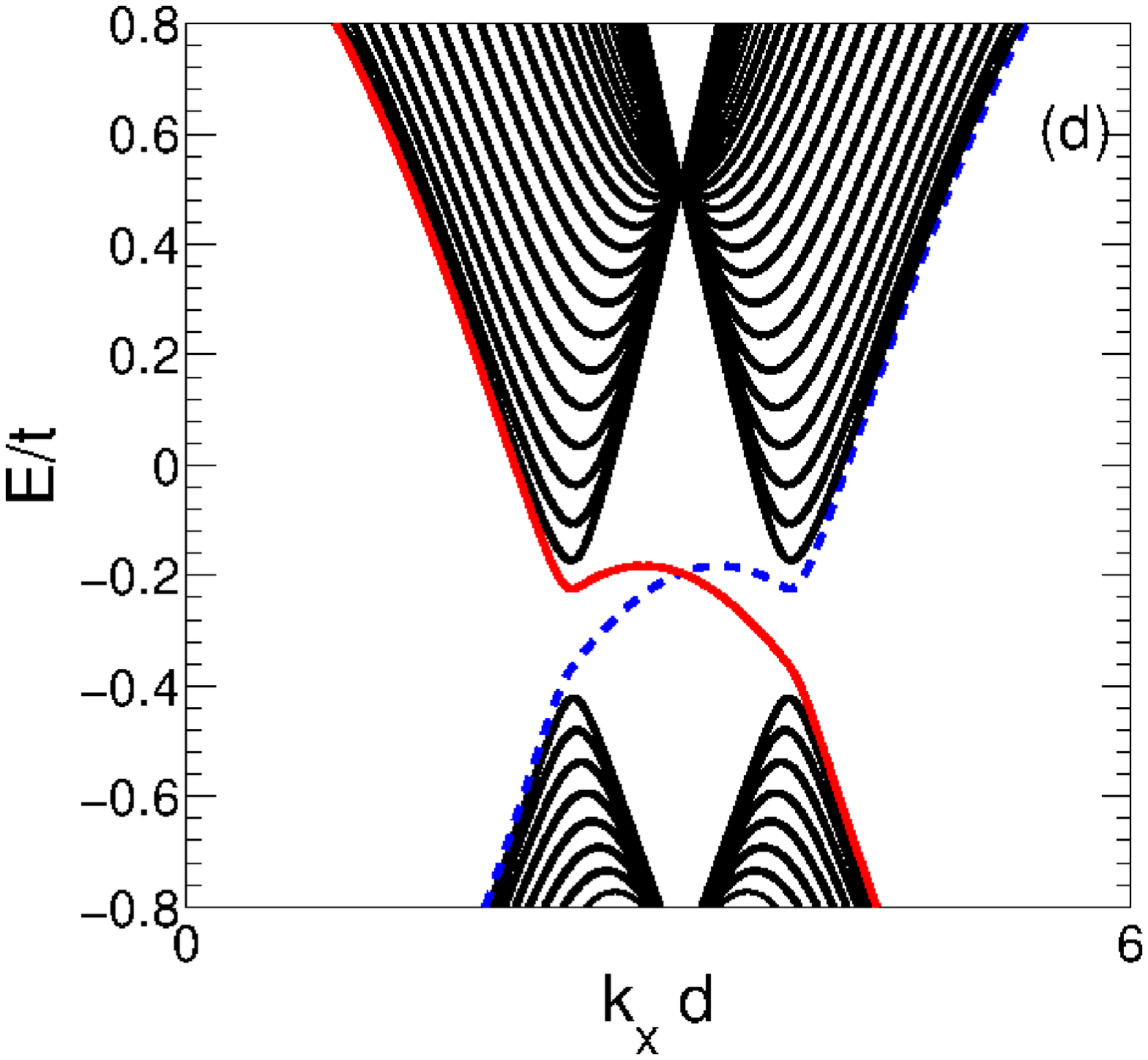}&
\includegraphics[width=0.5\columnwidth]{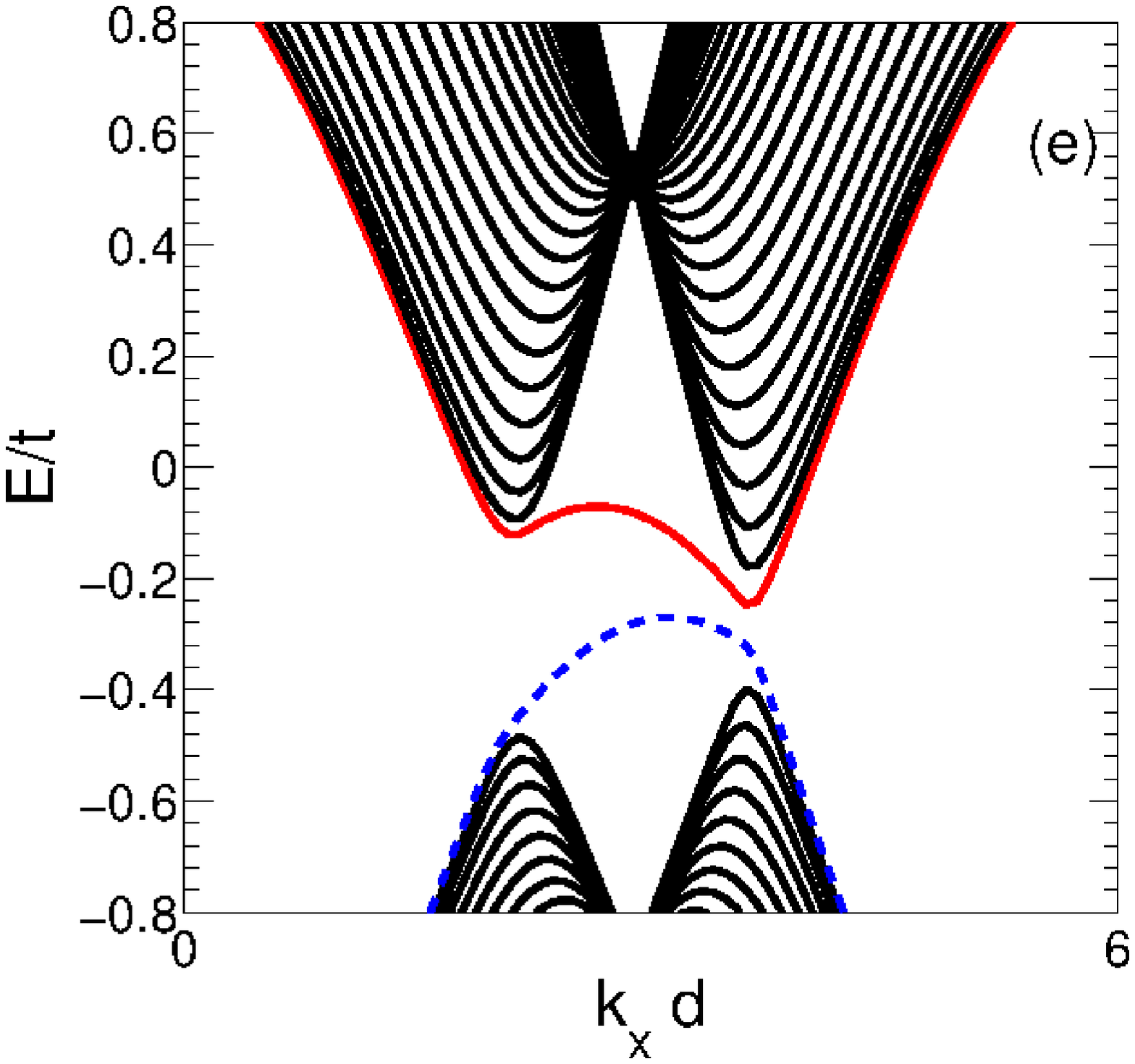}
\end{array}$
$\begin{array}{cc}
\includegraphics[width=0.5\columnwidth]{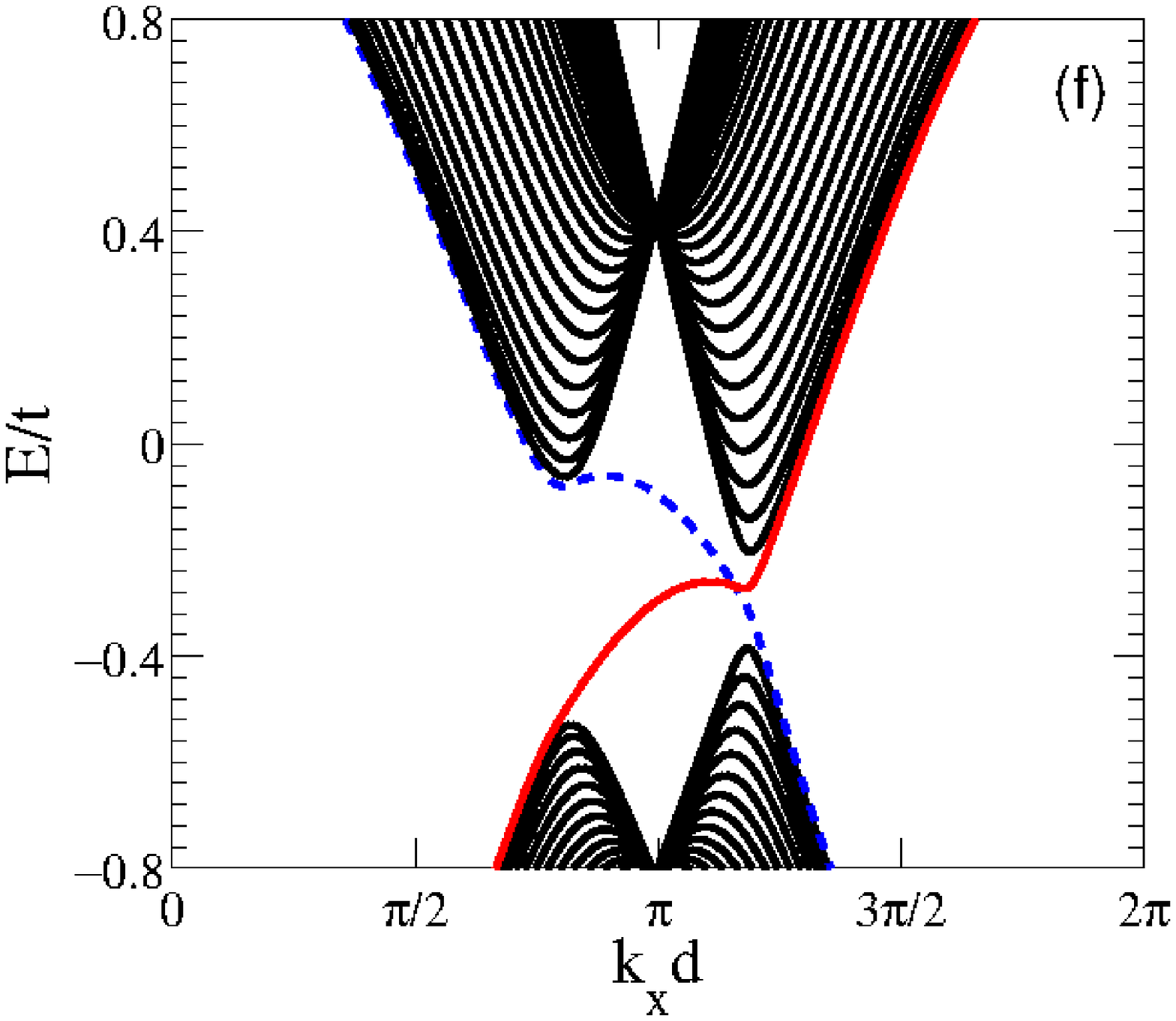}&
\includegraphics[width=0.5\columnwidth]{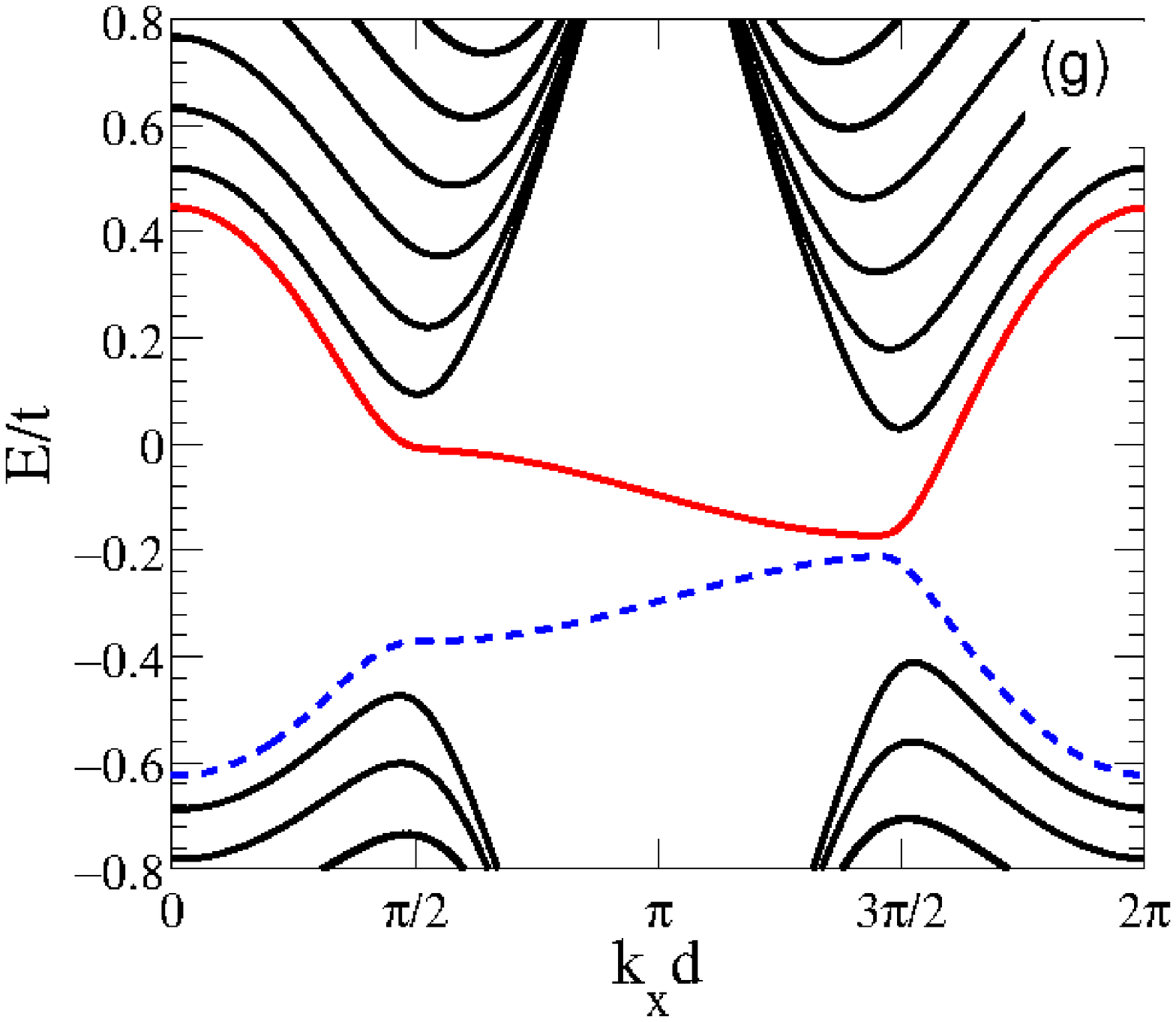}
\end{array}$
\end{center}
\caption{(a) Geometry of a zigzag nanoribbon of a width $W$ along the armchair edge. The unit cell is  shown by the dashed line. (b-e) Electronic band structure of HM within the tight binding approach for a zigzag nanoribbon with a width of $W=60$ atoms along the $y$ direction. For figures (b-d), calculations are done for $\Phi^{\prime}=(1+\epsilon)\Phi$ (Eq.\ref{phip}), $t_2=0.1 t$, $M=0$ and $\Phi=\frac{14}{15}\pi$. Figures (b), (c) and (d) correspond, respectively, to the undeformed lattice ($\epsilon=0$), $\epsilon=0.1$ at which the topological gap closes, and $\epsilon=0.15$. Figure (e) represents the case where $M=0.1t$, $\Phi=\frac{31}{36}\pi$ and $\epsilon=0.17$ ascribed to a trivial band insulator.
For the lower panels, we take $M=0.1t$,  $\Phi^{\prime}\Phi=\frac{14}{15}\pi$, $\epsilon=0.2$ (f) and $\epsilon=-0.2$ (g).}
\label{TB}
\end{figure}

We consider the HM in a strained zigzag nanoribbon deformed along the armchair direction. The ribbon geometry is shown in figure \ref{TB}(a). We calculate the full band structure within the TB model for a ribbon of a width $W=60$ atoms along the $y$ axis parallel to the strain direction. \

%\onecolumngrid

%\twocolumngrid

Figure \ref{TB} (b) represents the case of the HM on the undeformed lattice for a topological phase, with $M=0$ and $\Phi=\frac {14}{15}\pi$, for which the gap is purely topological and the Chern number is $\mathcal C=-1$. The gapless states crossing the gap are ascribed to the chiral edge states. The corresponding eigenfunctions show that these edge modes are localized on the bottom (solid red line) and on the top (blue dashed line) of the ribbon boundaries. By increasing the strain amplitude to $\epsilon_0=0.1$, the topological gap closes and the edge states become degenerate as in graphene zigzag nanoribbons, which corresponds to the phase boundary line in figure \ref{diagram}.\
For $\epsilon\sim 0.15$, the topological gap reopens and the chiral edge states reappear showing opposite slopes compared to the case $\epsilon<\epsilon_0$, which means that the corresponding edge currents will change signs (Fig.\ref{TB}(d)).
This result is consistent with the phase diagram of Fig.\ref{diagram} showing that, for $M=0$ and $\Phi=\frac {14}{15}\pi$, the Chern number changes from $\mathcal C=-1$, in the absence of strain, to $\mathcal C=1$ under a strain of $\epsilon\sim 0.15$.\\

Figure \ref{TB} (e) shows the behavior of the edge states under strain in the case where $M=0.1t$ and $\Phi=\frac{31}{36}\pi$. The topological phase ($\mathcal C=-1$) is tuned, at $\epsilon\sim 0.17$, to a trivial one ($\mathcal C=0$) for which the edge states disappear.\

Figures \ref{TB} (f) and (g) correspond to the case where $\Phi^{\prime}\Phi$. According to these figures, the strain can also tune the system from a topological phase (Fig.\ref{TB} (f)) to a trivial one (Fig.\ref{TB} (g)).

Following Reference \cite{Guassi}, we discuss the relationship between the dispersion of the edge states of a topological phase and the corresponding Chern number.

\begin{figure}[hpbt] 
\begin{center}
$\begin{array}{cc}
\includegraphics[width=0.5\columnwidth]{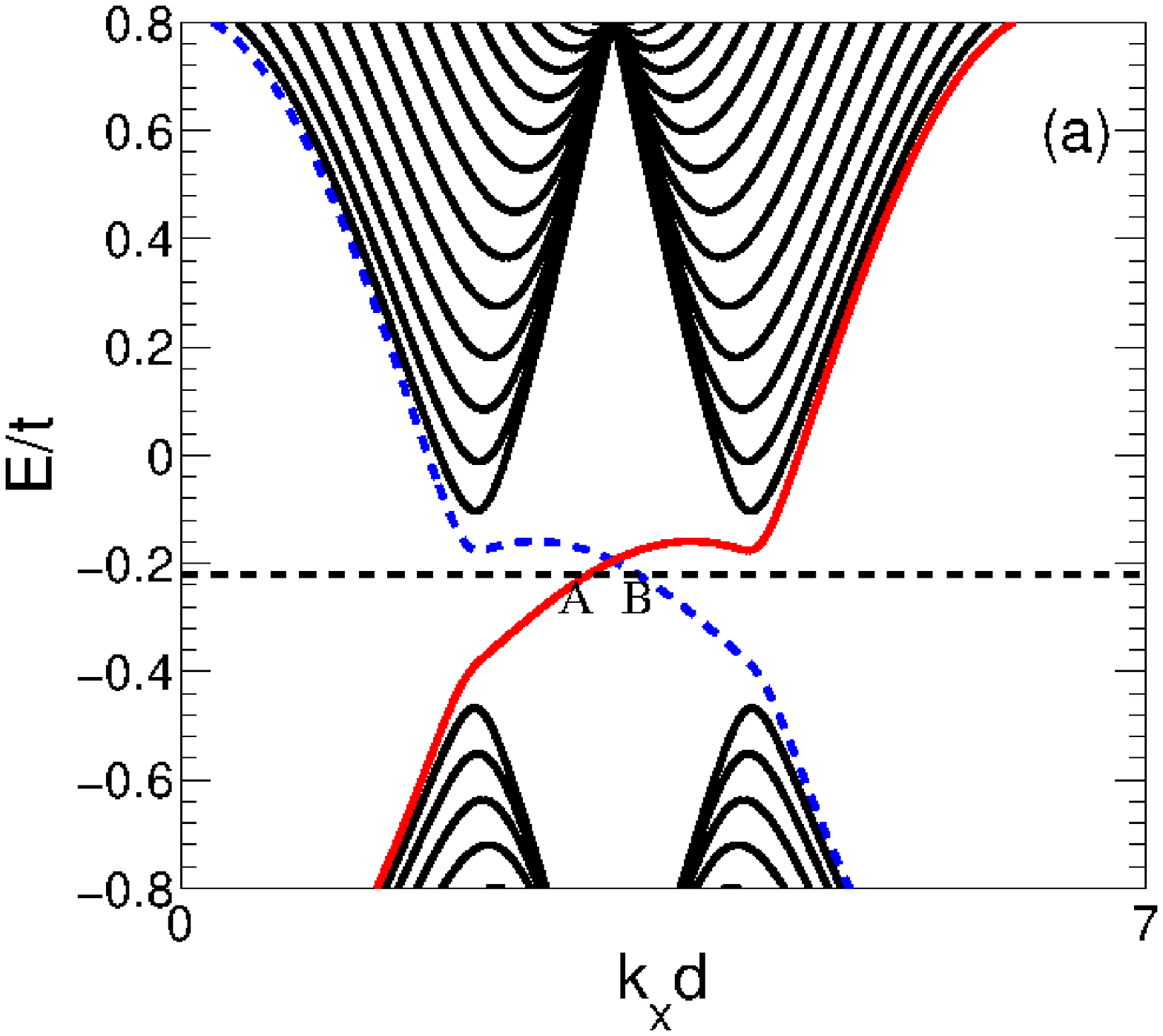}&
\includegraphics[width=0.5\columnwidth]{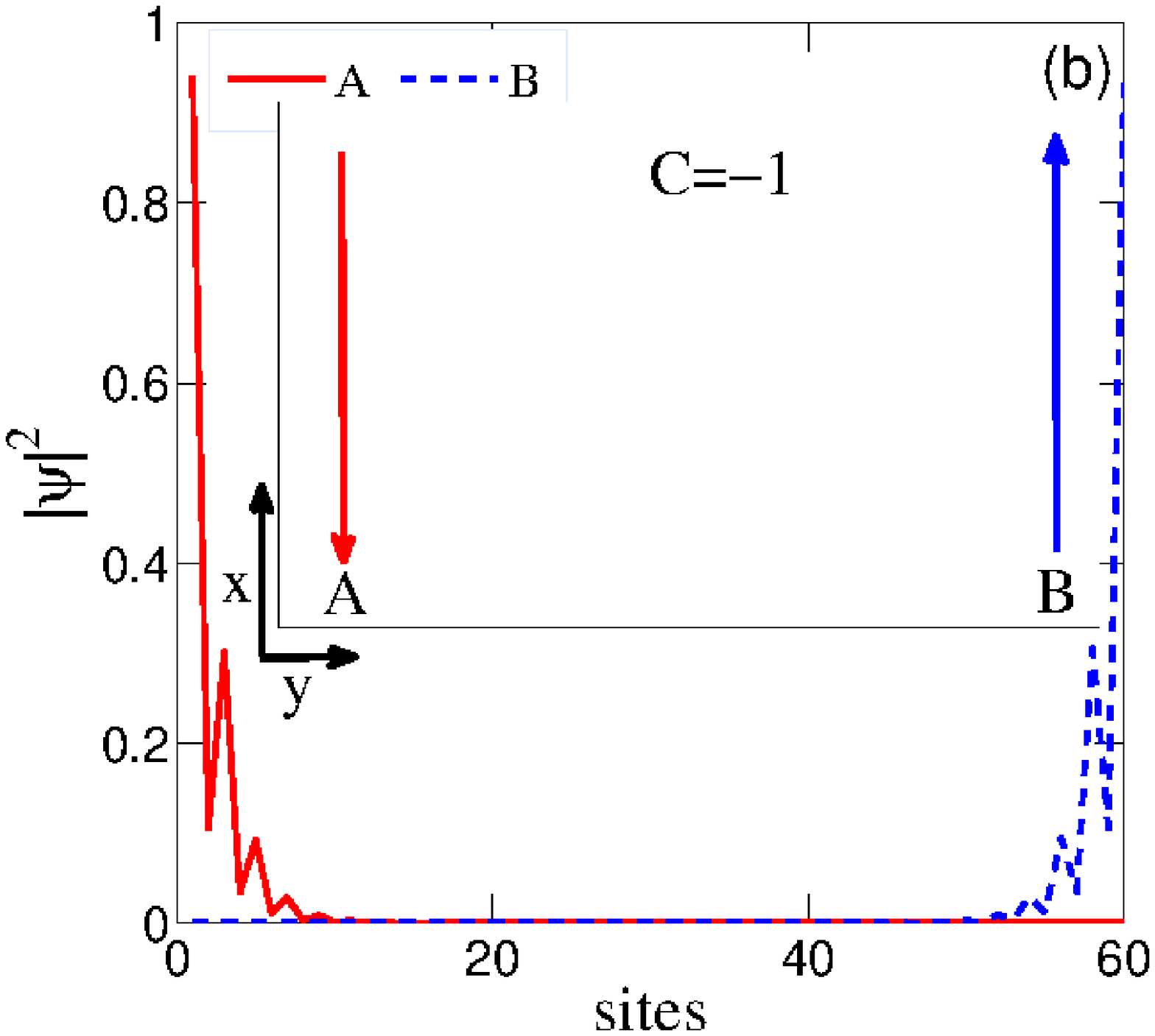}
\end{array}$
$\begin{array}{cc}
\includegraphics[width=0.5\columnwidth]{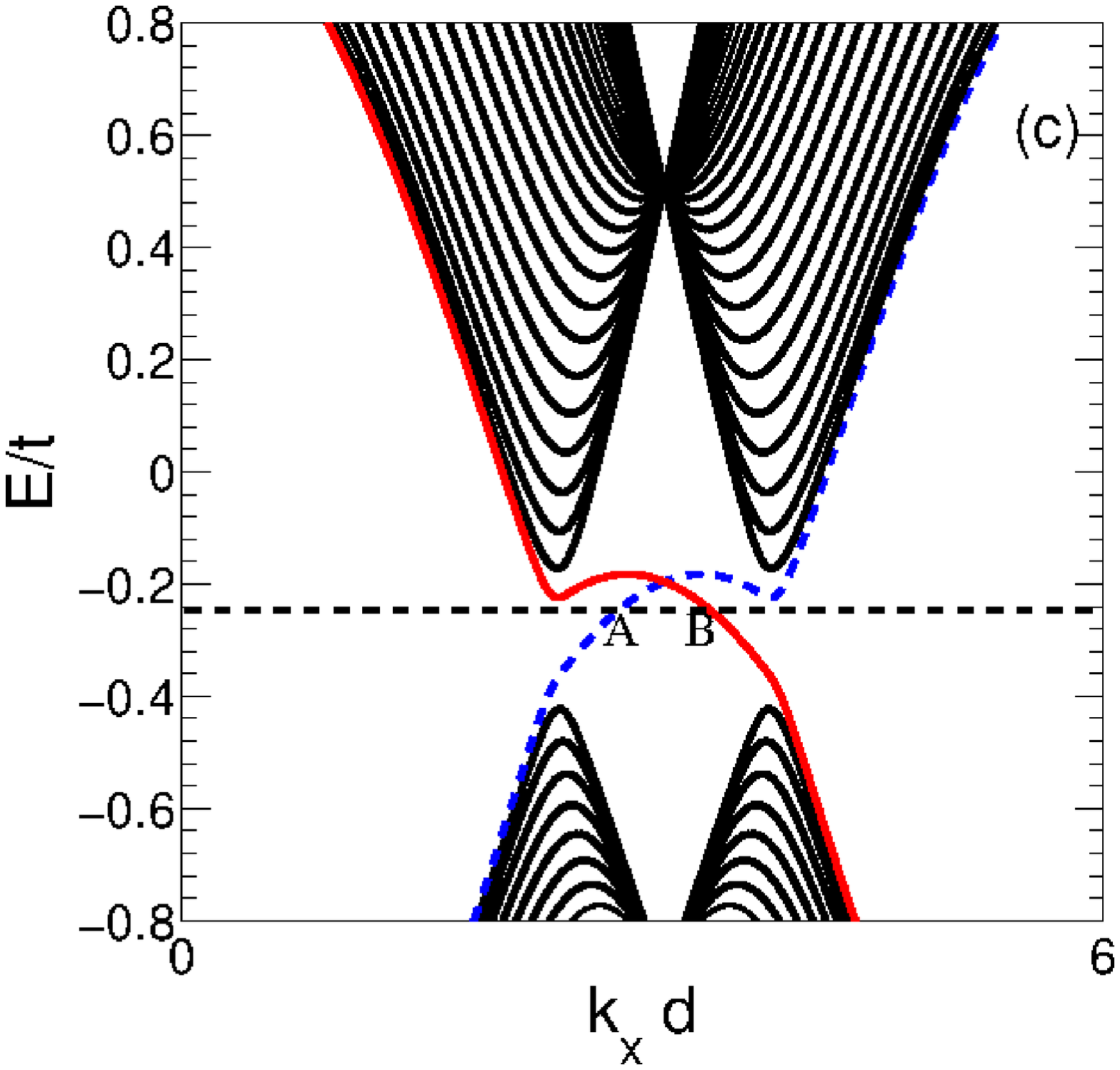}&
\includegraphics[width=0.5\columnwidth]{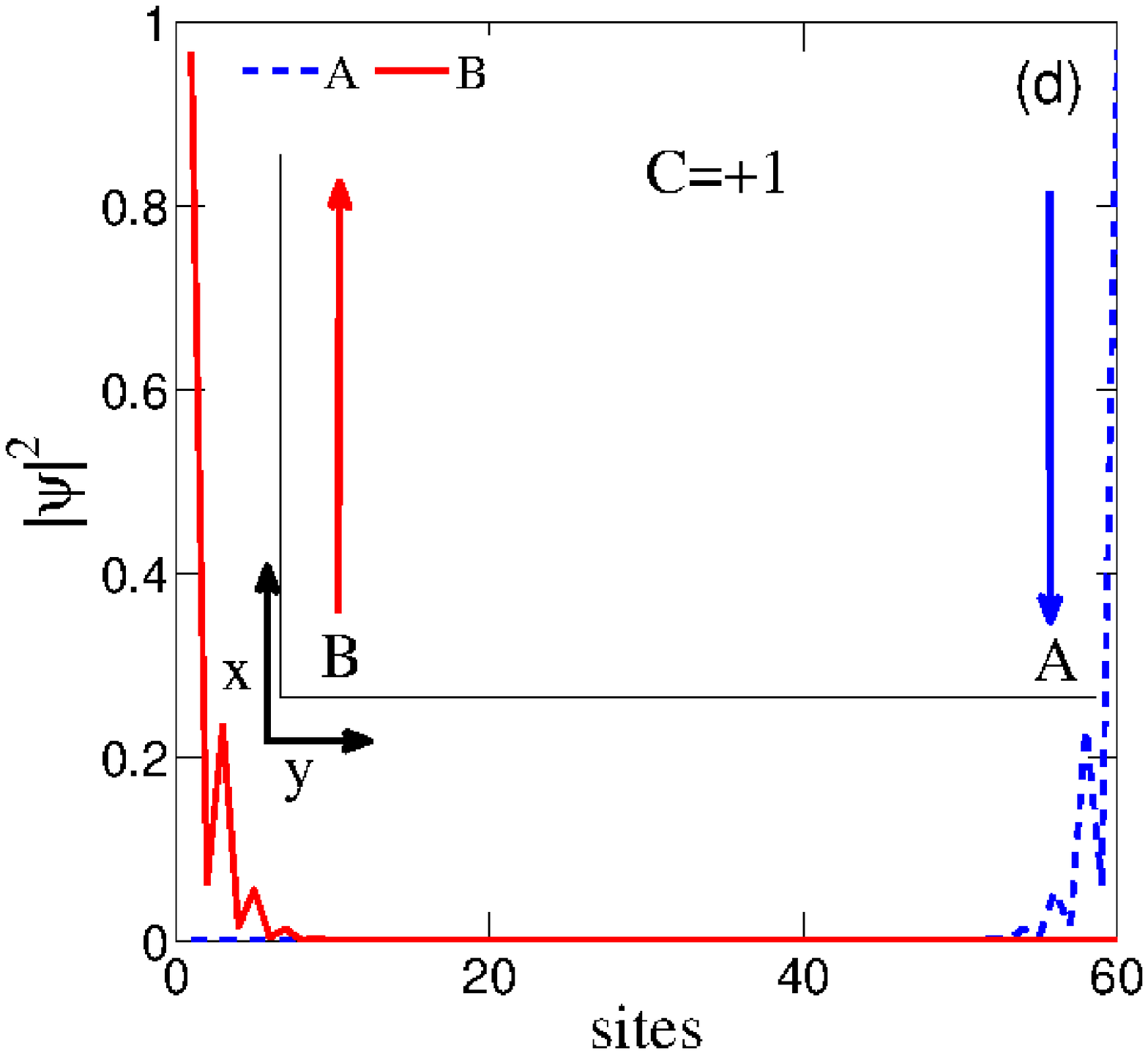}
\end{array}$
\end{center}
\caption{Probability distributions of the edge states of a zigzag nanoribbon of a width $W=60$ atoms described by the HM under a strain amplitude of $\epsilon=0$ (upper panels) and $\epsilon=0.15$ (lower panels) and for $M=0$ and $\phi=\frac{14}{15}\pi$. The figures (a) and (b) ((c) and (d)) correspond to a topological phase with a Chern number $C=-1$ ($C=1$). The insets in figures (b) and (d) give the direction of the edge currents. Calculations are done for $\Phi^{\prime}=(1+\epsilon)\Phi$ (Eq.\ref{phip}).}
\label{proba}
\end{figure}

Figure \ref{proba} represents the energy spectrum of the underformed system for $M=0$ and $\Phi=\frac {14}{15}\pi$, which corresponds, according to the Haldane phase diagram to a topological phase with $\mathcal C=-1$.
The probability distributions of the edge states, denoted $A$ and $B$, are represented in Fig.\ref{proba} (b), which shows that the $A$ ($B$) edge state with the positive (negative) velocity $v_x=\frac 1{\hbar} \frac{\partial E(\vec{k})}{\partial k_x}$ is localized on the bottom (top) boundary of the ribbon.\
For a Fermi level above the zero energy, the edge states will give rise to edge currents $I=-ev_x$, $e>0$ being the elementary charge. These currents, depicted in Figs.\ref{proba} (b,d), are responsible of the sign of the Chern number since the anomalous Hall conductivity reads as: $\sigma_{xy}=\frac{e^2}{h}\mathcal C$.
In Fig.\ref{proba} (b) ((d)), the current is negative (positive) which yields to $\mathcal C=-1$ ($\mathcal C=+1$), in agreement with the phase diagram obtained in the continuum limit (Fig.\ref{diagram}).

\section{Modified Haldane model under uniaxial strain}

\subsection{Electronic Hamiltonian}

The mHM was first studied by Varney et al.\cite{mHM} who showed that this model could exhibit non zero circulating edge currents when the chiral symmetry is broken. In this case, the symmetry properties of the system are reminiscent of the nonquantized anomalous Hall effect.\newline
Colomés and Franz \cite{Colomes} have reported that a strip described by mHM can hold antichiral edge modes propagating in the same direction and compensated by bulk modes.\

Recently, the optical absorption properties of the mHM have been discussed in Ref.\onlinecite{Vila} where the authors predicted the possibility to realize simultaneously circular dichroism and valley polarization, which may pave the way to applications combining light polarization and valleytronics effects.\

Antichiral edge states were also predicted to occur in exciton-polariton honeycomb lattice on ribbon with zigzag edges \cite{Mandal}. These photonic antichiral edge states are expected to be of a great interest in information processing regarding their robustness against disorder.\

\begin{figure}[hpbt] 
\begin{center}
\includegraphics[width=0.4\columnwidth]{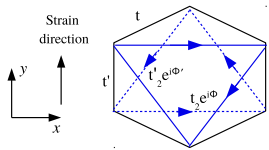}
\end{center}
\caption{The pattern for the second-neighbor hopping parameters of the mHM under uniaxial strain applied along the armchair direction. The arrow indicate the directions along which the hopping integrals $t_2$ and $t_2^{\prime}$ acquire positive phase $e^{i\Phi}$ and $e^{i\Phi^{\prime}}$ respectively.}
\label{mHM}
\end{figure}

 Using the same approach discussed in Sec.II, one can derive the continuum limit of the electronic dispersion relation for the mHM where the hopping integrals to the second neighboring atoms are given by the pattern shown in figure \ref{mHM}.

\begin{figure}[hpbt] 
\begin{center}
$\begin{array}{ccc}
\includegraphics[width=0.3\columnwidth]{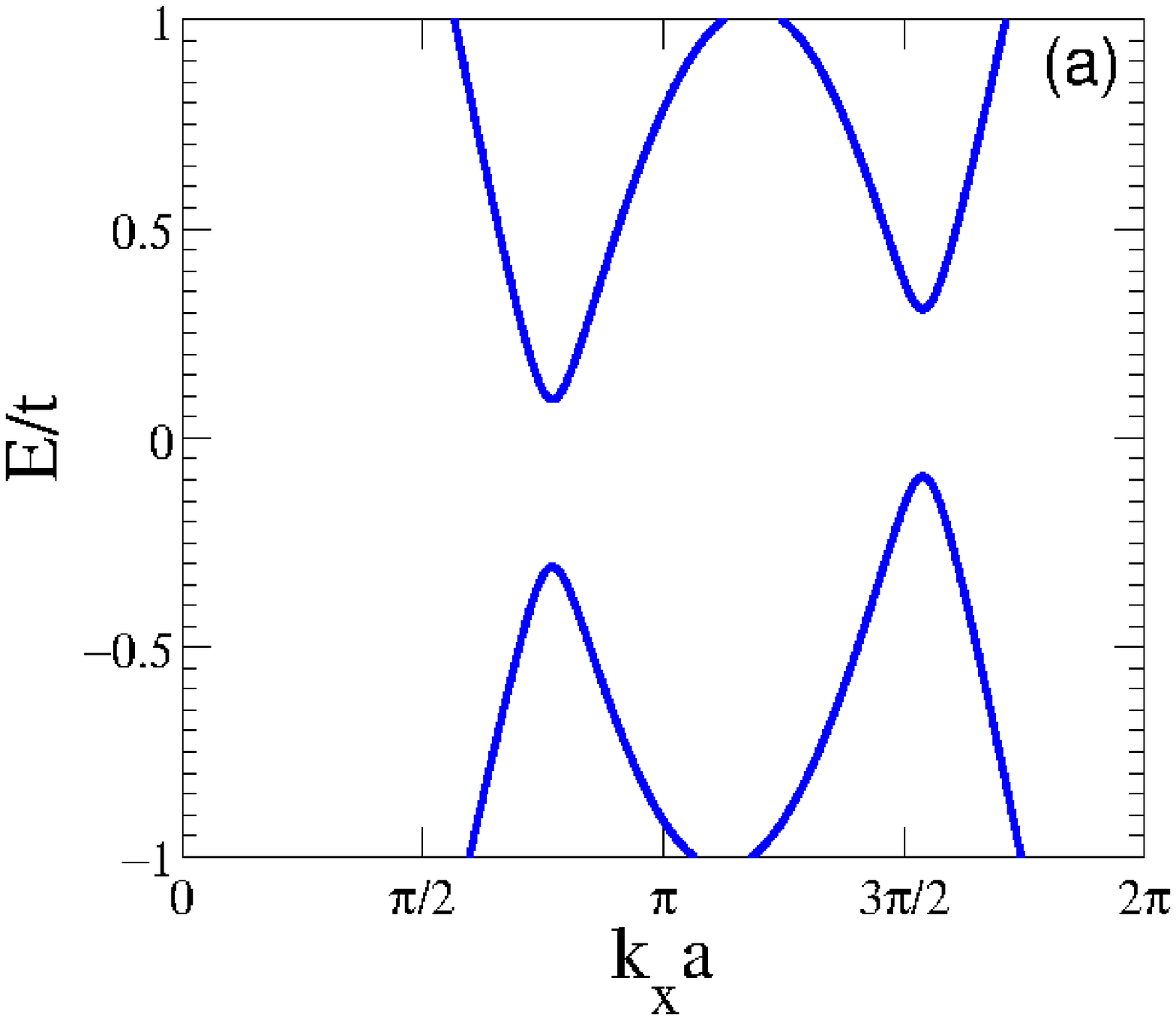}&
\includegraphics[width=0.3\columnwidth]{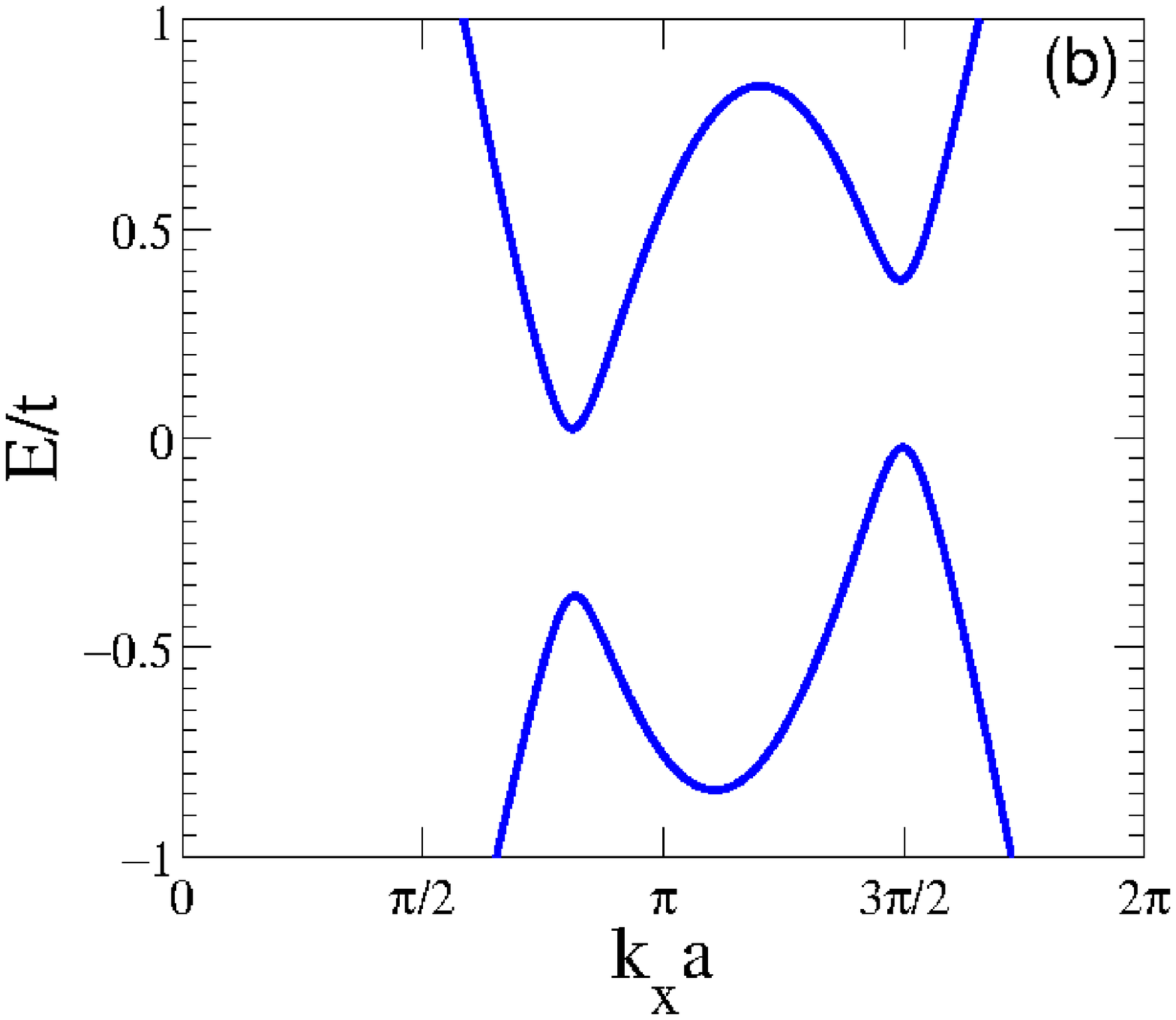}&
\includegraphics[width=0.3\columnwidth]{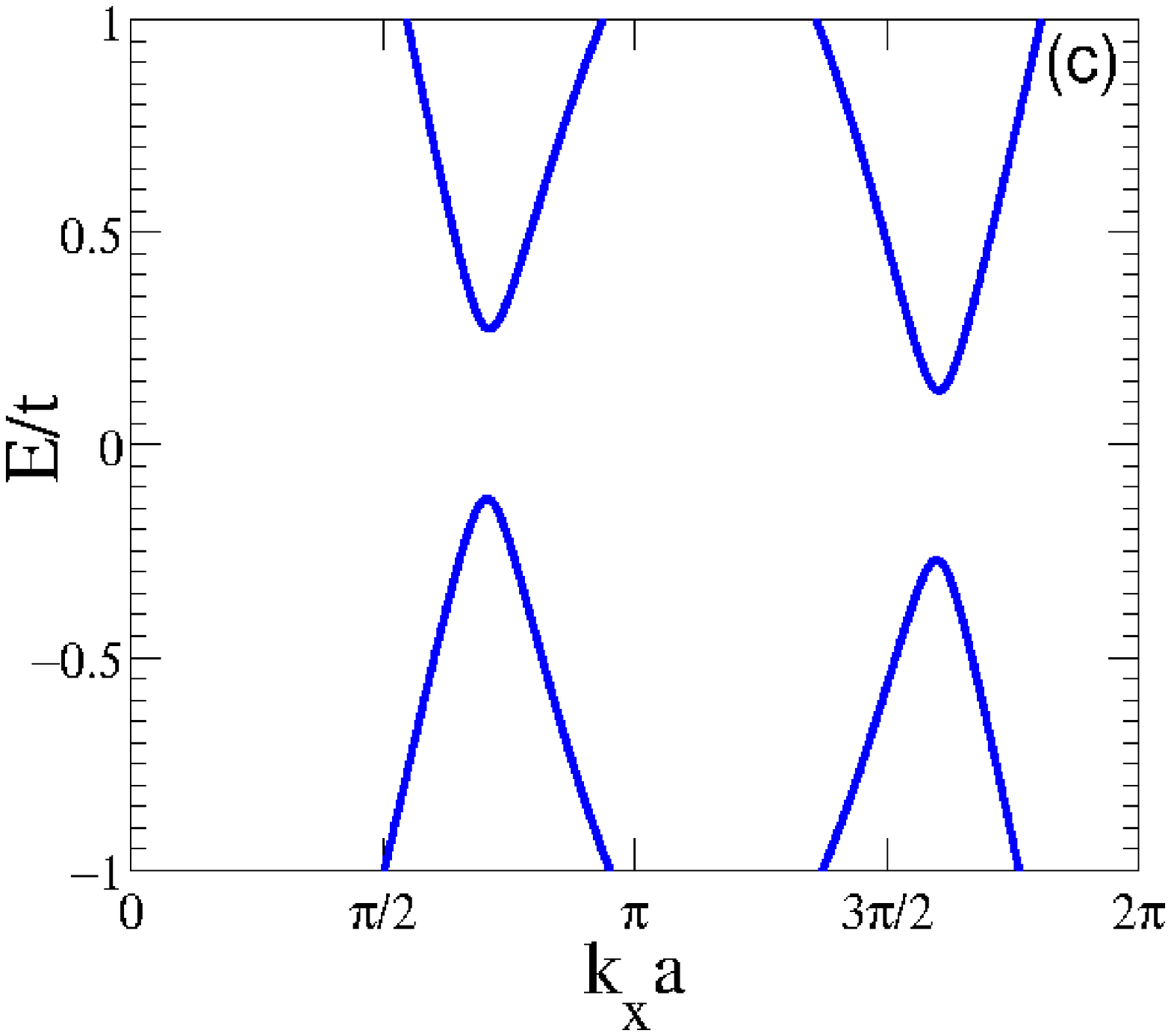}
\end{array}$
$\begin{array}{ccc}
\includegraphics[width=0.3\columnwidth]{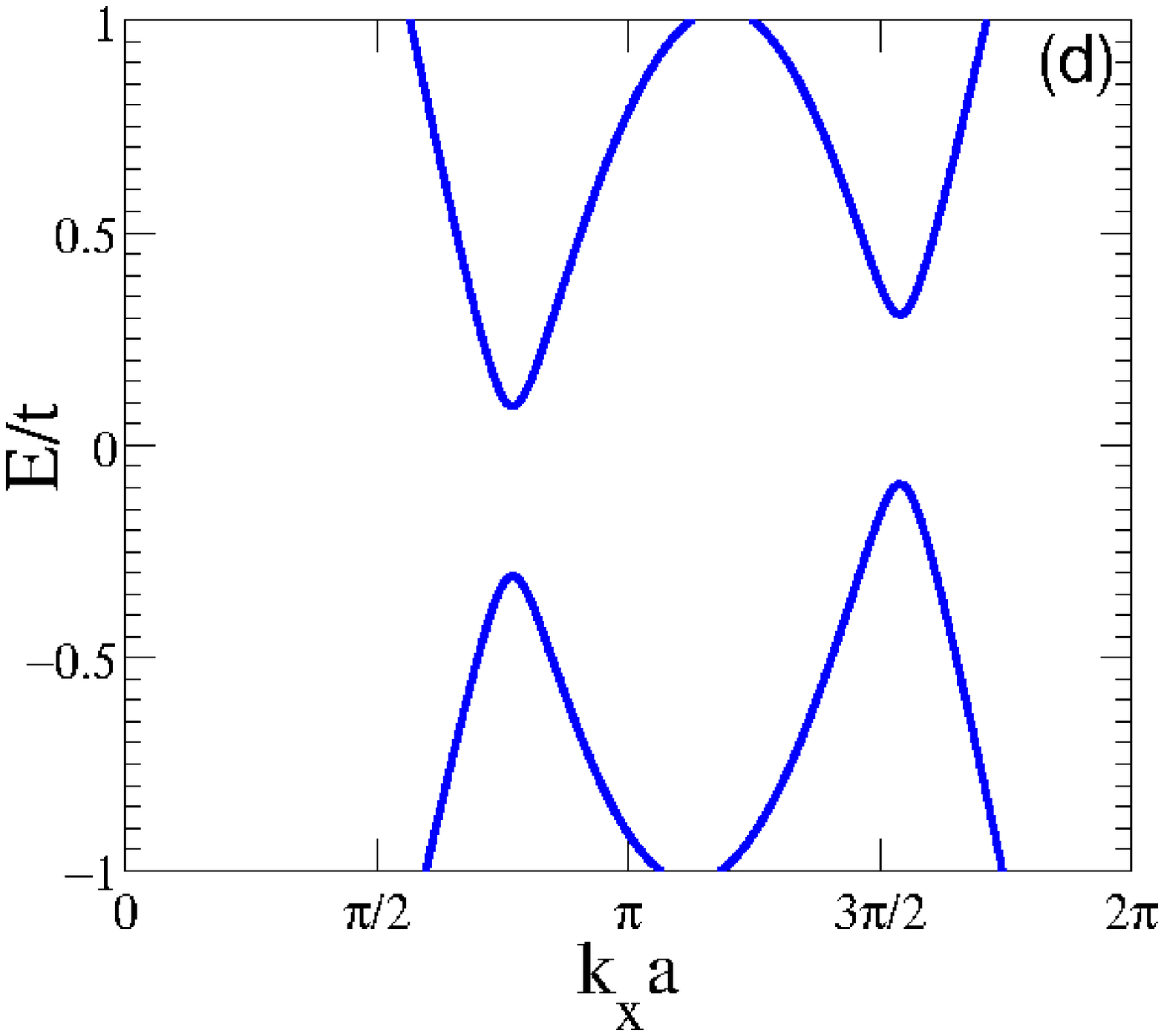}&
\includegraphics[width=0.3\columnwidth]{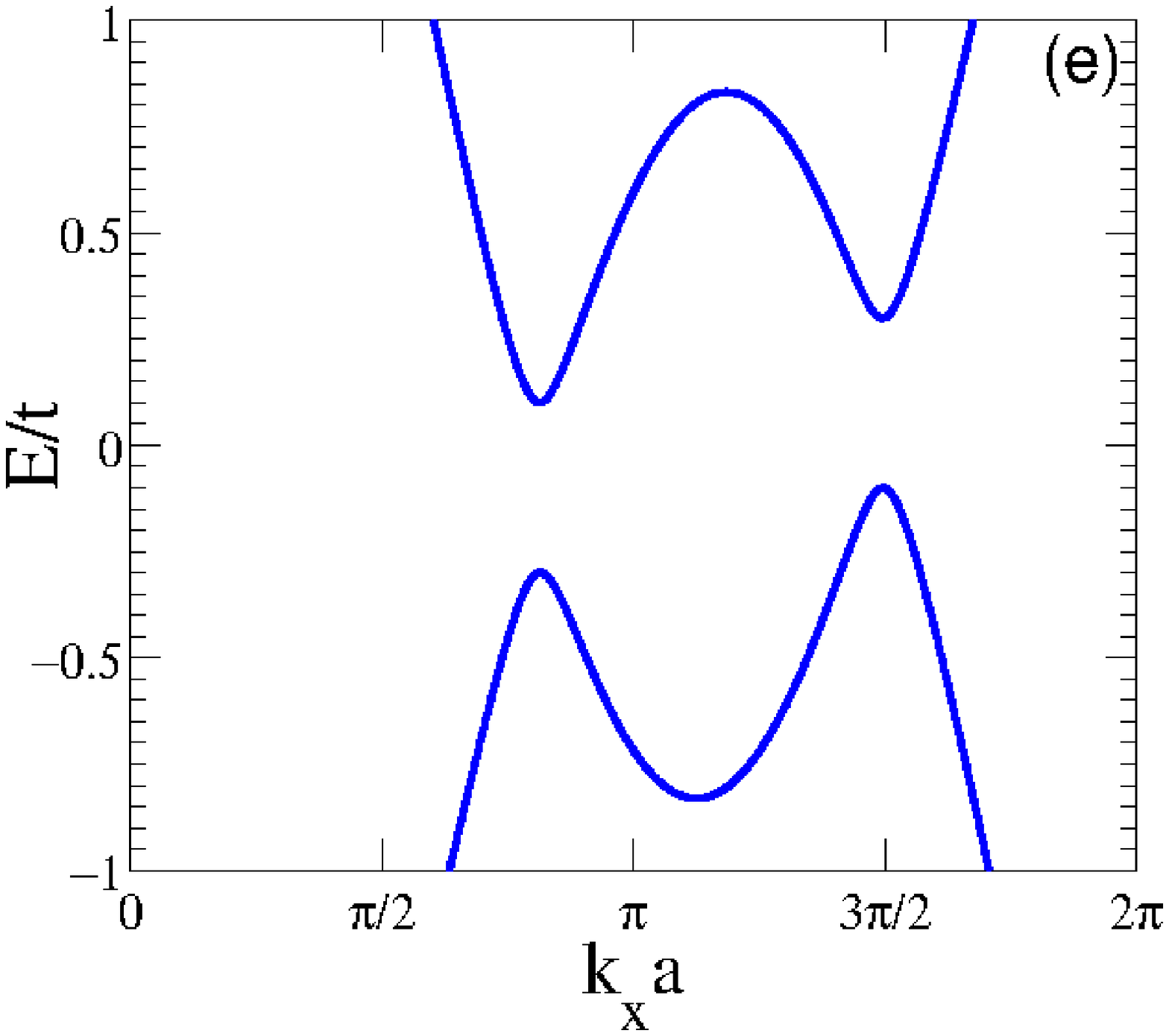}&
\includegraphics[width=0.3\columnwidth]{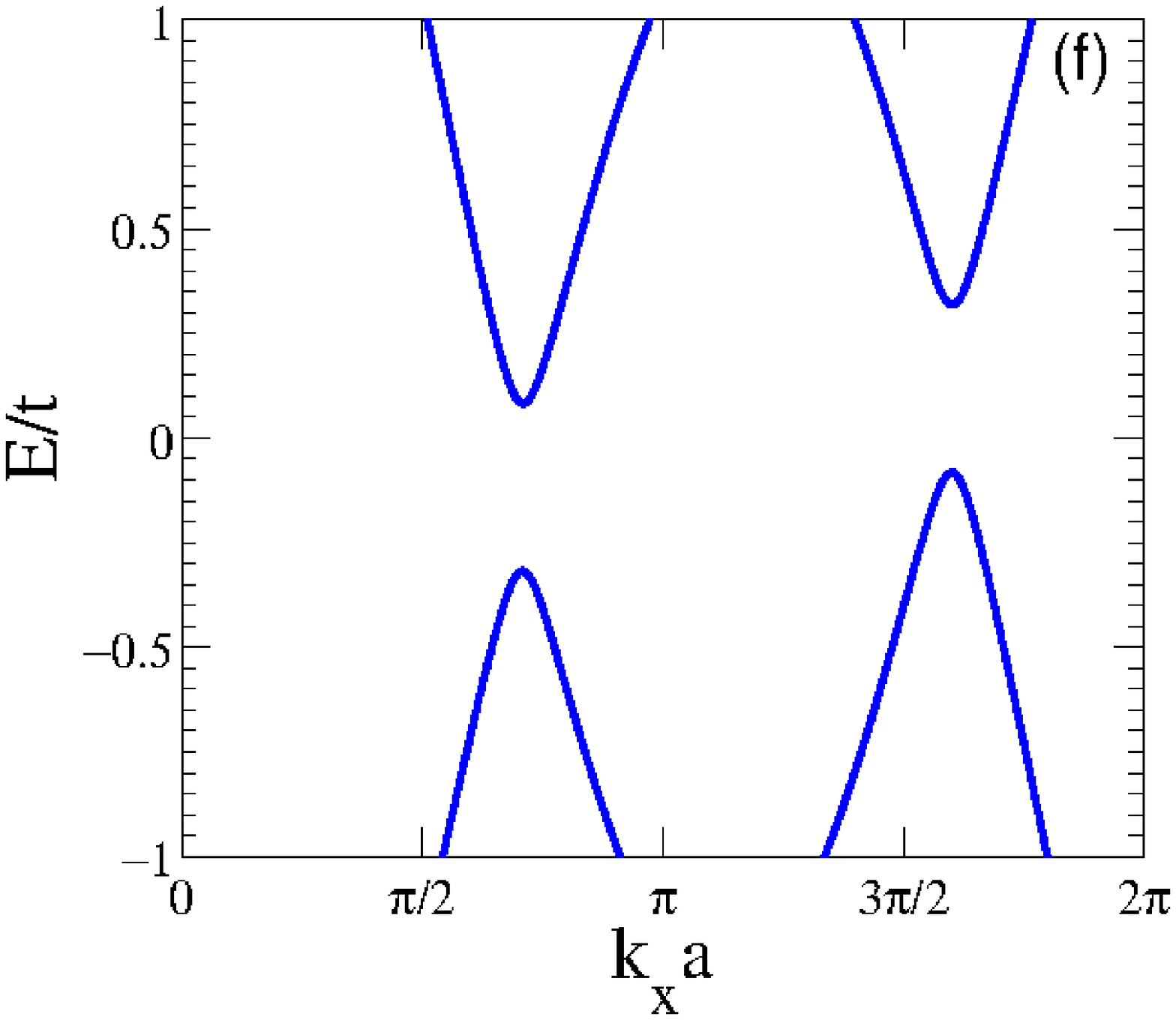}
\end{array}$
\end{center}
\caption{Electronic band structure at low energy of the mHM for $t_2=0.1t$, $k_y=0$, and for the Semenoff mass $M=0.2t$.
The upper panels correspond to $\Phi=\frac{14}{15}\pi$ and $\Phi^{\prime}$ given by Eq.\ref{phip} while the lower ones are for $\Phi=\Phi^{\prime}$. The strain amplitude is  $\epsilon=0$ (a) and (d), $\epsilon=-0.1$(b) and (e) and $\epsilon=0.15$ (c) and (f).}
\label{mHMenergy}
\end{figure}

The diagonal term $h_{AA}$ of the Hamiltonian given by Eq.\ref{H2x2} is the same as in the HM. However, the $h_{BB}$ (Eq.\ref{hbb}) is changed since now, one should replace $\Phi$ and $\Phi^{\prime}$ by, respectively, $-\Phi$ and $-\Phi^{\prime}$.
The diagonal terms of the Hamiltonian given by Eq.\ref{Hgeneral} become:

\begin{eqnarray}
h_{0}(\vec{k})&=&2t_2\left(\cos \Phi \cos\vec{k}.\vec{a}_1+\frac{t_2^{\prime}}{t_2}\cos\Phi^{\prime}(\cos\vec{k}.\vec{a}_2+\cos\vec{k}.\vec{a}_3)\right) \nonumber\\
&&-2t_2\left(\sin \Phi \sin\vec{k}.\vec{a}_1+\frac{t_2^{\prime}}{t_2}\sin\Phi^{\prime}(\sin\vec{k}.\vec{a}_2-\sin\vec{k}.\vec{a}_3)\right)\nonumber \\
h_{z}(\vec{k})&=& M 
\end{eqnarray}
The corresponding low energy dispersion relation takes the form:

\begin{equation}
\epsilon_{\lambda}^{\xi}(\vec{q})=\xi\left(m_0+\hbar w_{0x}q_x\right)+\lambda\hbar \sqrt{w_{x}^2 q_x^2+ w_{y}^2q_y^2+M^2},
\label{disper2}
\end{equation}
where the mass term $m_{0}$ is :
\begin{equation}
 m_{0}=2t_2\left(\frac{2t^{\prime}_2}{t_2}\sin\Phi^{\prime} \sin\theta -\sin\Phi \sin 2\theta\right)
 \label{mass0}
\end{equation}
In figure \ref{mHMenergy} we depicted the low energy dispersion relation for different strain values in the cases where 
$\Phi=\Phi^{\prime}$ and $\Phi\neq\Phi^{\prime}$ ($\Phi^{\prime}$ given by Eq.\ref{phip}).

\subsection{Modified Haldane model: tight binding approach}

We consider a graphene nanoribbon with zigzag edges under a uniform uniaxial strain applied along the armchair direction (denoted $y$ axis). The ribbon has a finite width $W$ along the armchair edge. This system could exhibit co-propagating-edge states as shown in Ref.[\onlinecite{Colomes}].\

\begin{figure}[hpbt] 
\begin{center}
$\begin{array}{cc}
\includegraphics[width=0.4\columnwidth]{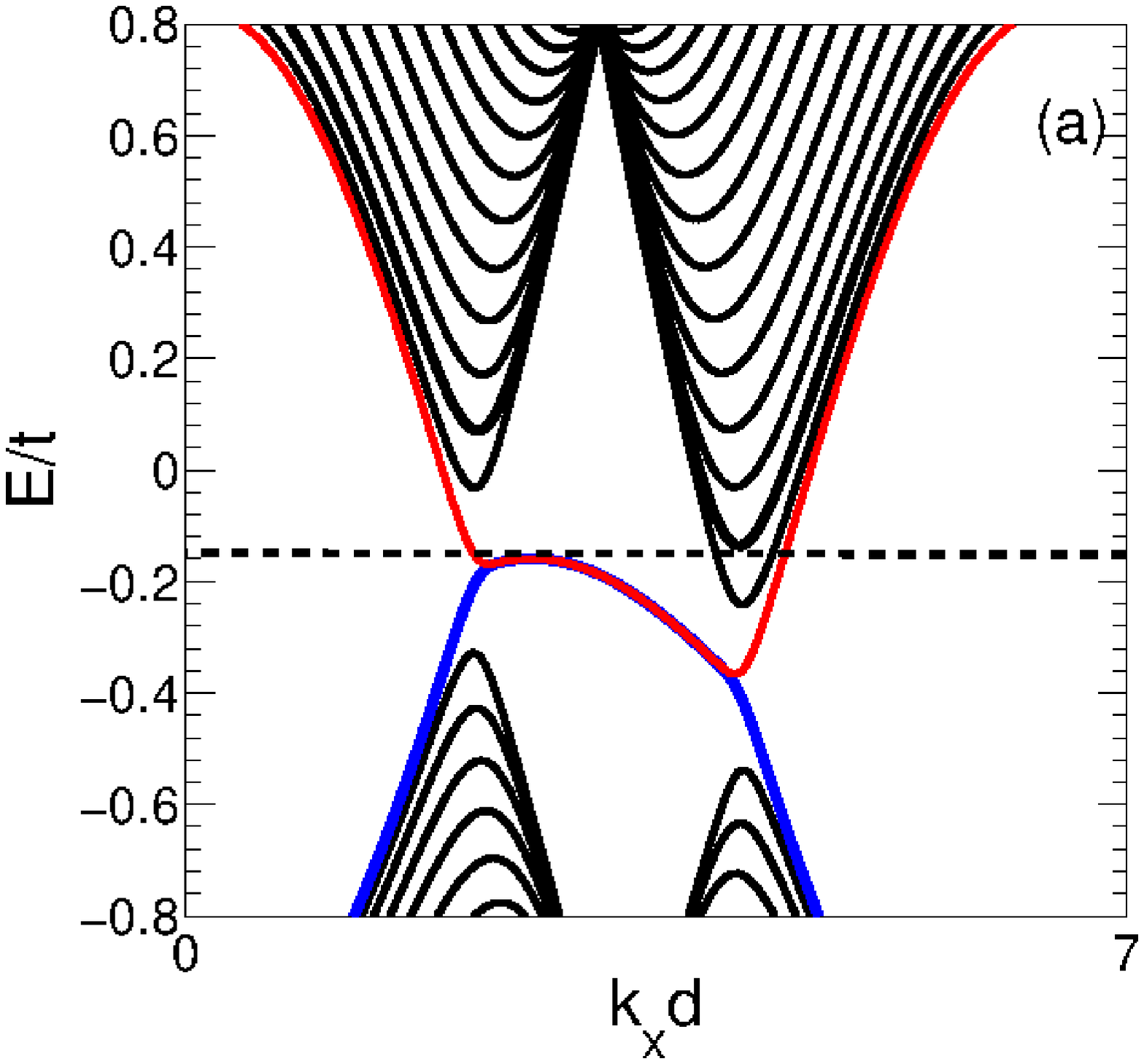}&
\includegraphics[width=0.4\columnwidth]{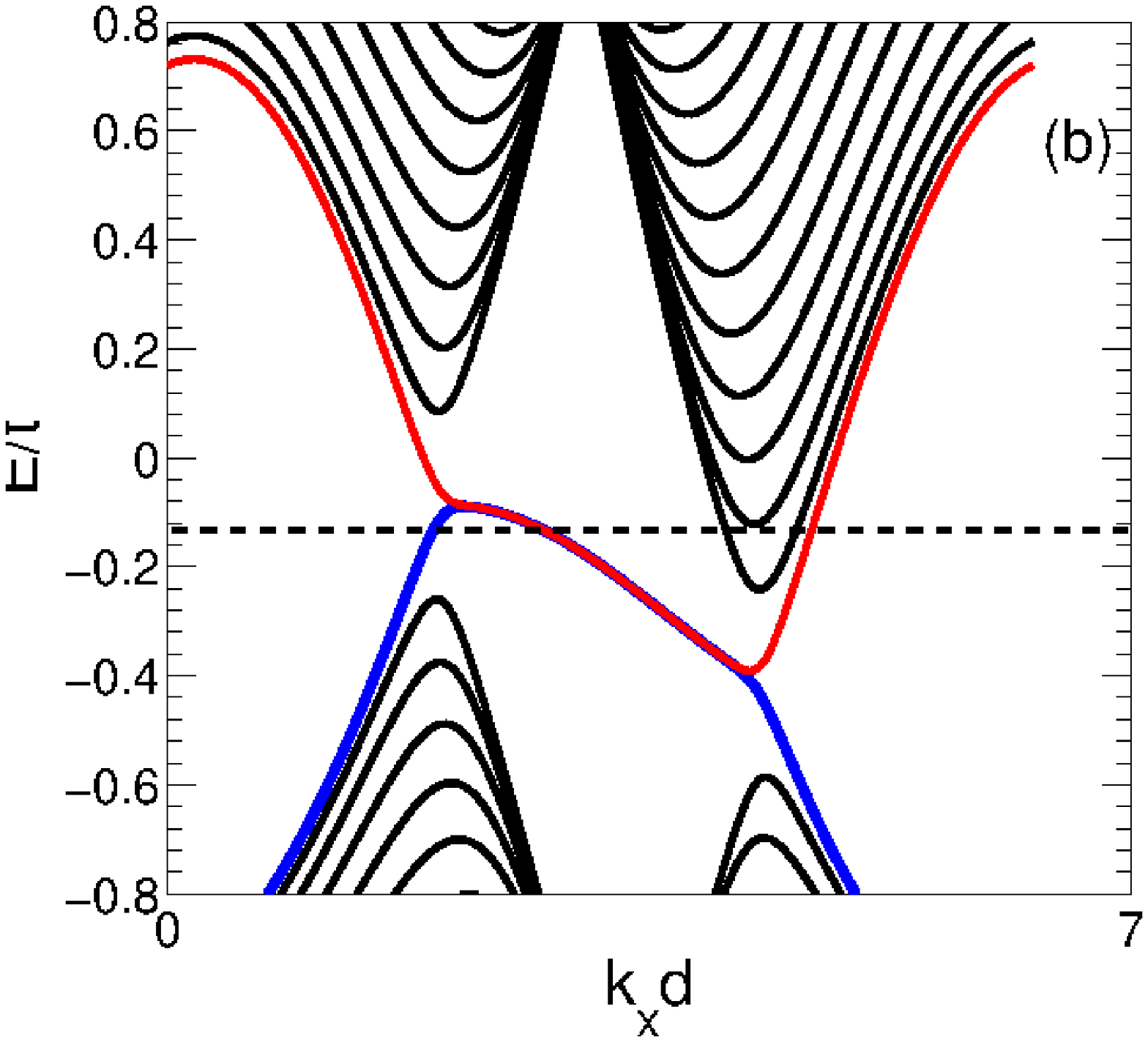}
\end{array}$
$\begin{array}{cc}
\includegraphics[width=0.4\columnwidth]{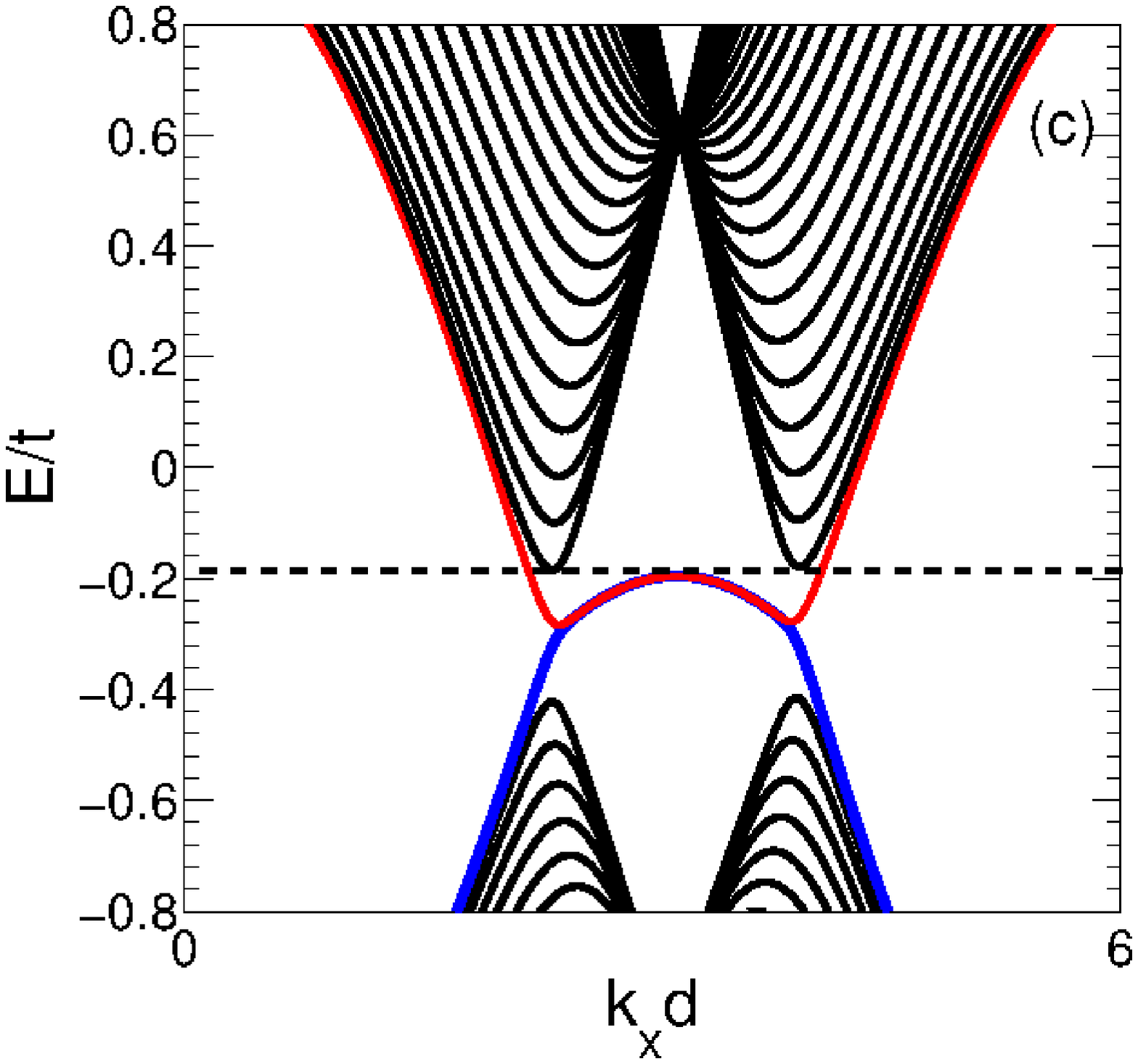}&
\includegraphics[width=0.4\columnwidth]{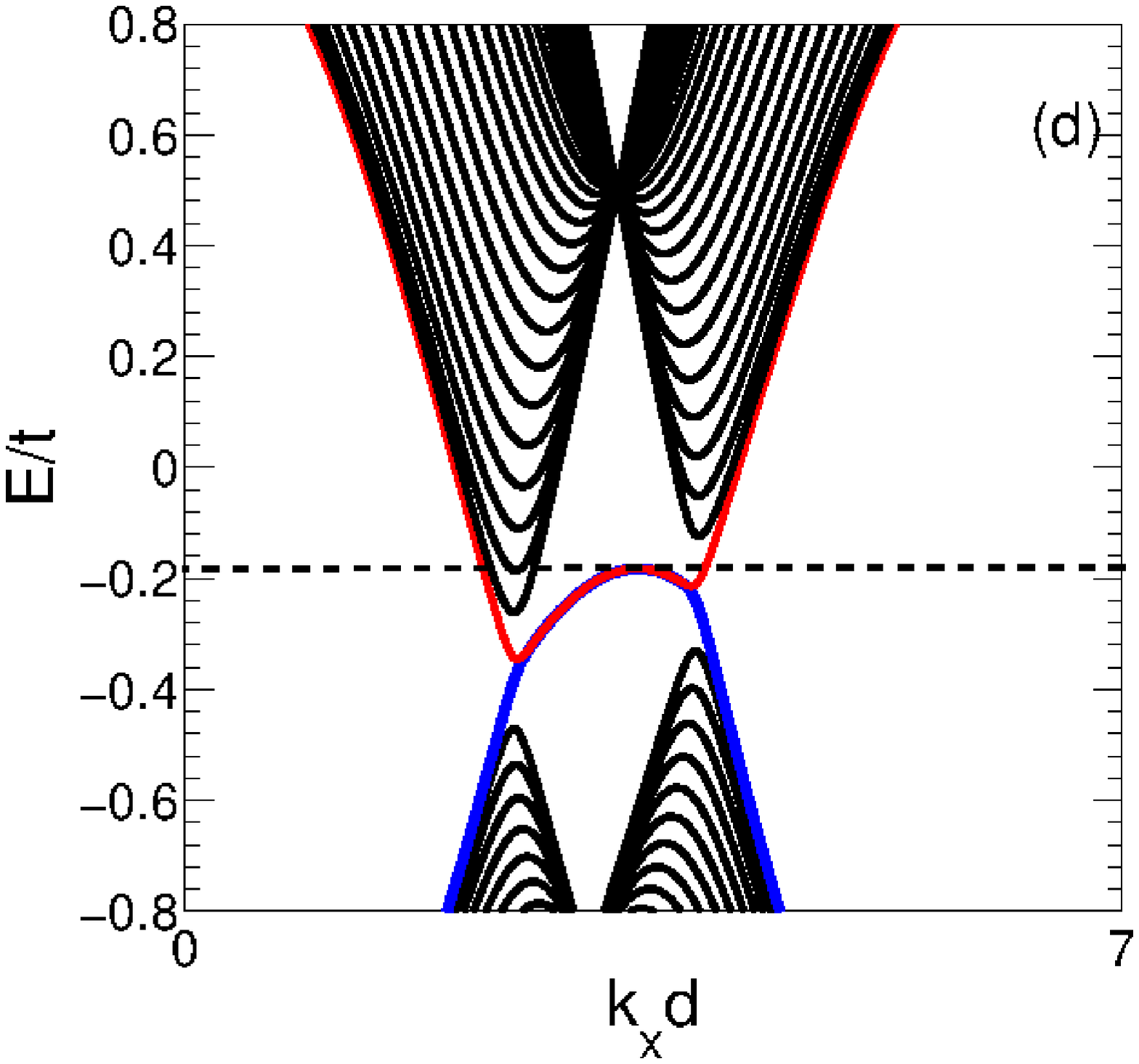}
\end{array}$
$\begin{array}{cc}
\includegraphics[width=0.4\columnwidth]{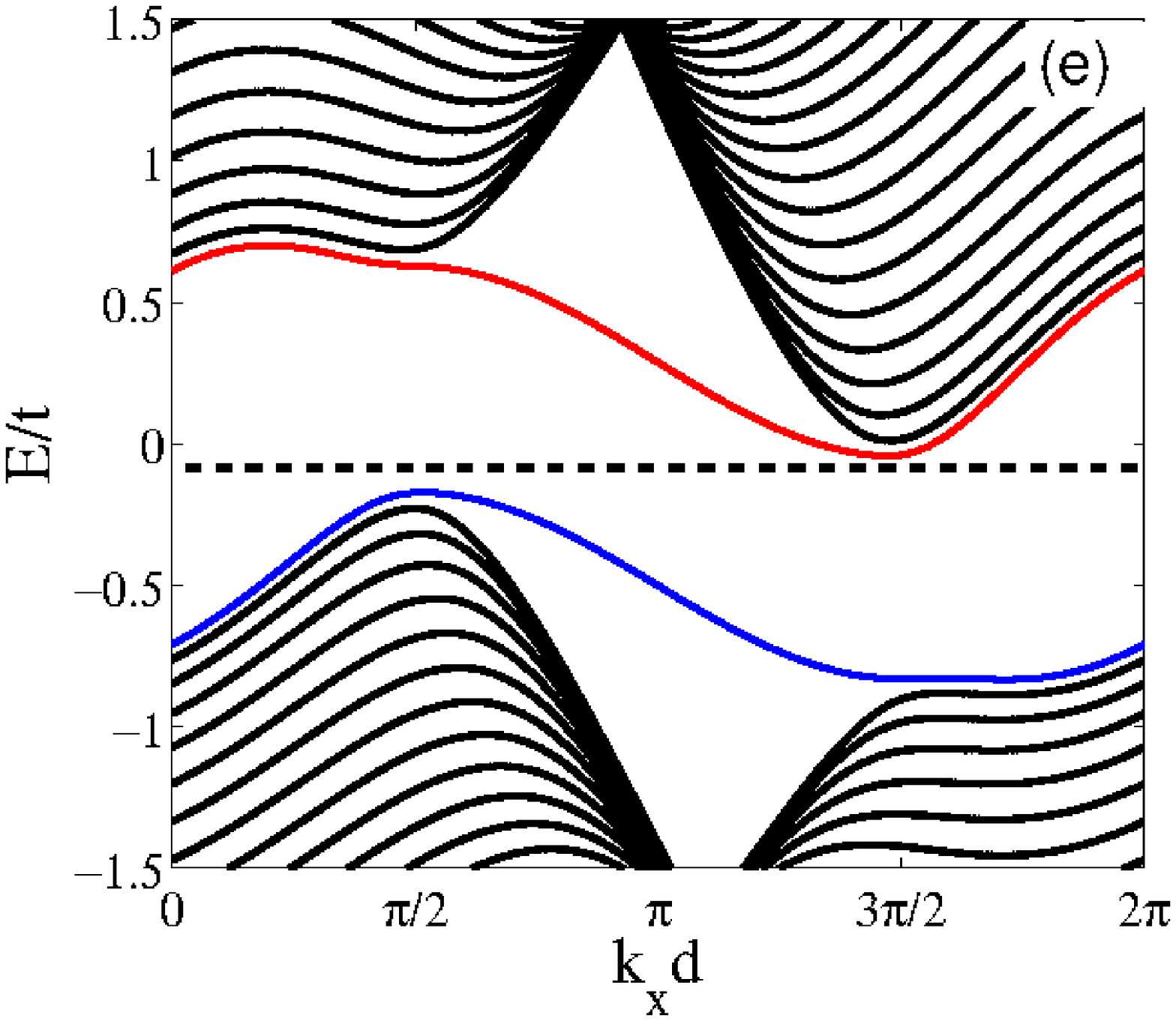}&
\includegraphics[width=0.4\columnwidth]{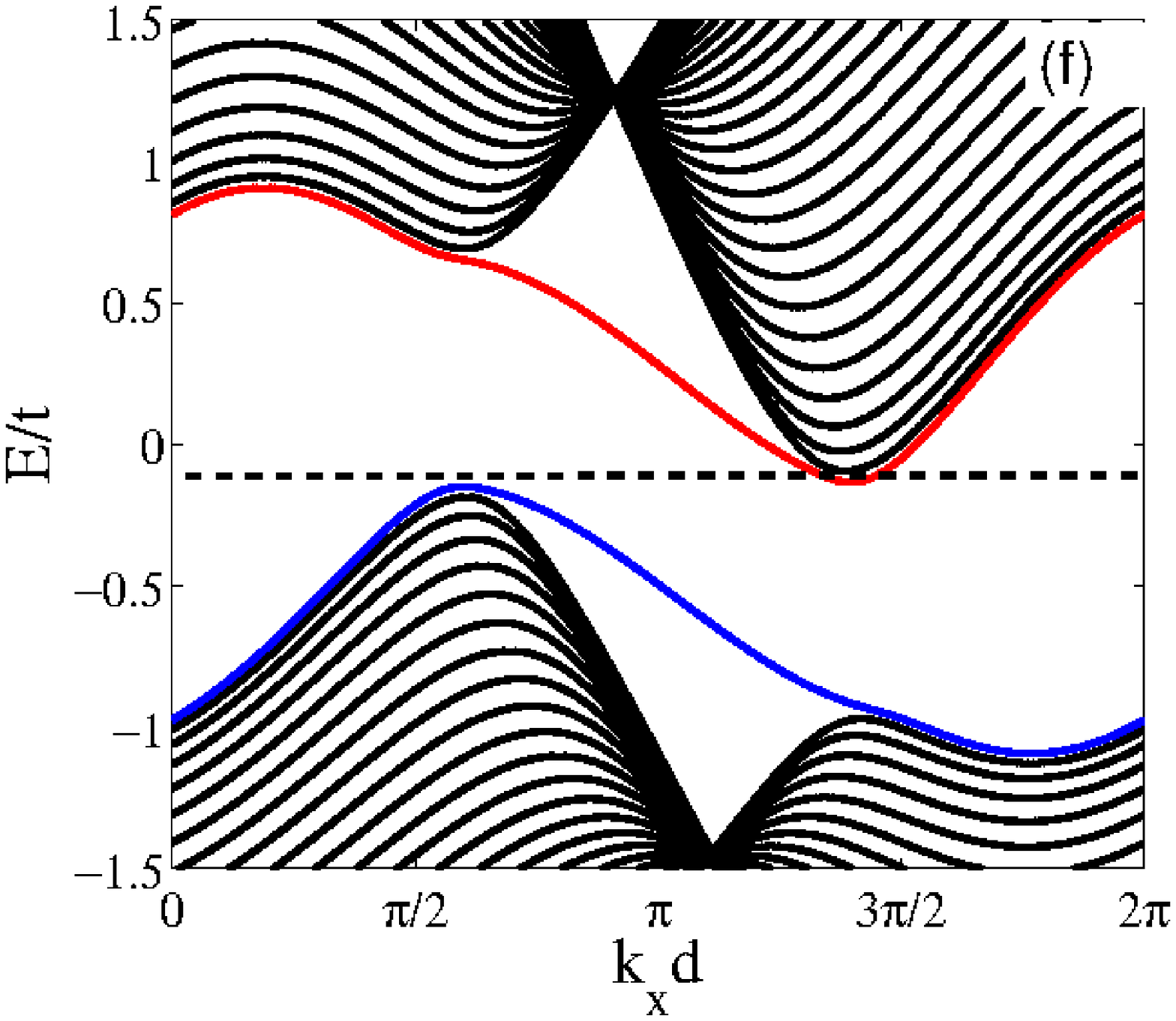}
\end{array}$
\includegraphics[width=0.4\columnwidth]{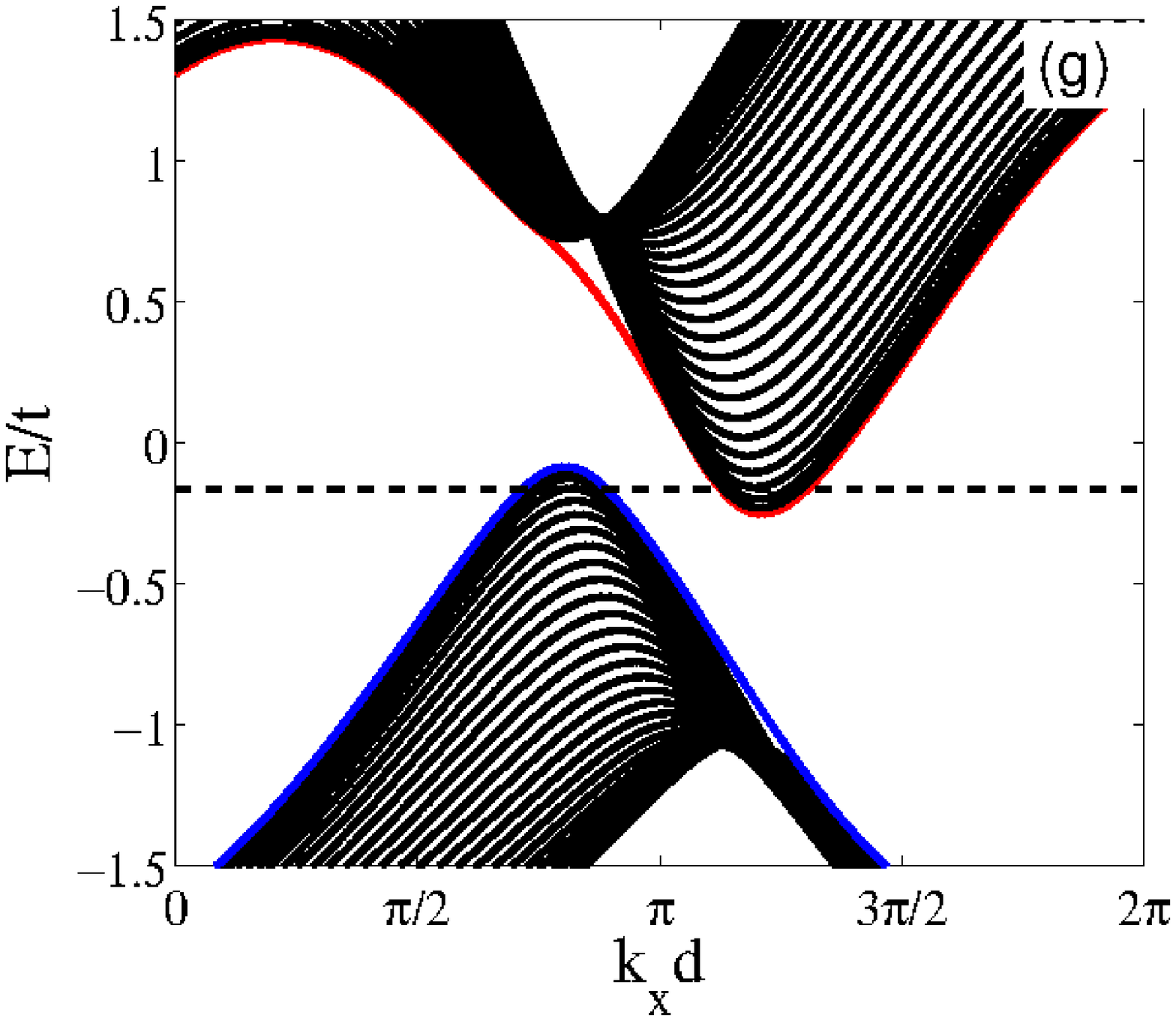}
\end{center}
\caption{Band structure of a graphene zigzag nanoribbon under uniaxial strain described by the mHM at different strain amplitudes. The calculations are done for a ribbon of a width of $W=60$ atoms and $t_2=0.1t$. 
The panels (a-d) are for $M=0$, $\Phi=\frac{14}{15}\pi$ and $\Phi^{\prime}=(1+\epsilon)\Phi$ and correspond, respectively, to the undeformed lattice ($\epsilon=0$), $\epsilon=-0.08$, $\epsilon=\epsilon_0=0.1$ and $\epsilon=0.15$. At the critical value of $\epsilon_0=0.1$, the dispersion of the edge modes is flatten and above this value their velocity changes sign compared to the case $\epsilon<\epsilon_0$. The dashed line indicates the position of the Fermi level. 
for the lower panels (e-g), calculations are done for $M=0.4t$, $\Phi^{\prime}=\Phi=0.68\pi$ and $\epsilon=-0.25$ (e),  
$\epsilon=-0.1$ (f), at which the mass term $m_0$ (Eq.\ref{mass0}) vanishes and $\epsilon=0.2$ (g).
}
\label{mHM-energy}
\end{figure}

%%%
Using the tight binding approach, we depict in figure \ref{mHM-energy} the electronic band structure of a ribbon of width $W=60$ atoms, at different strain amplitudes, for $M=0$ and $\Phi=\frac{14}{15}\pi$ as in the HM.
Figure \ref{mHM-energy} (a) shows that, in the undeformed lattice, the antichiral edge states have the same velocity which is counterbalanced by the bulk mode crossed by the Fermi energy.\

By increasing the strain amplitude, the dispersion of the antichiral edge modes is modified, and beyond a critical strain value $\epsilon_0\sim 0.1$, the antichiral edge states acquire opposite velocity compared to the case where $\epsilon<\epsilon_0$.\

The critical value $\epsilon_0$ corresponds to $m_0=0$ (Eq.\ref{mass0}) where the Dirac points are at the same energy leading to a dispersionless edge states. In the case where $\Phi^{\prime}$ obeys to Eq.\ref{phip}, and for small strain amplitudes ($|\epsilon| \ll1$), we find $\epsilon_0=-3\frac{\tan\Phi}{2\Phi}$, which is consistent with the numerical results depicted in figure \ref{mHM-energy}(c). For $\Phi^{\prime}=\Phi$, the energy offset of the Dirac points vanishes at the critical value  
$\epsilon_0=0.5$ (Fig.\ref{mHM-energy}(f)).
This feature may open the way to realize strain-tuned antichiral edge currents, which can be tested in the 2D metal transition dichalcogenide material $\mathrm{WSe_2}$. \newline
It is worth to note that the single layer structure of this material has a topological phase ($1T^{\prime}$ phase) and a trivial semiconducting one with an hexagonal structure ($1H$ phase)\cite{Chen}. As mentioned by Colomés and Franz \cite{Colomes}, by an appropriate doping, the $1H$ phase of WSe2 could support antichiral edge current of the valence band if the corresponding edge state energy is crossed by the Fermi level. We, then, expect to realize the strain modified antichiral edge states in the deformed $1H$-WSe2.
Mandal {\it et al.}\cite{Mandal} have, recently, proposed to use photonic systems to obtain antichiral edge states. To mimic the role of the strain, one can modify the graphene like geometry of the polariton strip, considered in Ref.\onlinecite{Mandal}, to induce a change in the dispersion of the antichiral edge modes.
We propose that the strained mHM could be realized in deformed 2D material WSe$_2$, doped in a way that the Fermi level crosses the edge state of the valence band, as proposed in Ref.\onlinecite{Colomes}. The edge current, for a given spin orientation, could be tuned by the strain. A strained WSe$_2$ has been recently used to achieve a single-photon emitter, a building block for quantum computing devices\cite{Rosenberger,Liu}.

\section{Conclusions} We discussed the robustness of the topological phases of the Haldane and the modified Haldane models against a uniform uniaxial strain. We considered a zigzag hexagonal nanoribbon exhibiting dispersionless edge states in the trivial phase and in the absence of strain. Using the continuum limit approximation and the tight binding approach, we found that the topology of these models could be tuned by the strain. At a critical value of the strain amplitude, a topological phase can be turned into a trivial one. By varying the strain amplitude, transitions between phases, with different Chern numbers, could take place. Our results show that the line boundary of the Haldane phase diagram, where one valley becomes gapless, is strain dependent. Moreover, the $2\pi$ periodicity of this line is lost in the case where the magnetic fluxes become strain dependent and the phase diagram shows a pseudo-periodicity which increases (decreases) for a compressive (tensile) deformation. Such behavior  may be probed by optical lattices of cold atoms.
We also showed that the dispersion of the topologically protected edge modes in the HM could be modified by the strain. The directions of propagation of the latter and the corresponding Chern number signs may be reversed by the strain. This feature may be realized in Fe-based ferromagnetic insulators \cite{Kee} or in graphene with SOC doped with magnetic atoms \cite{Guassi}.\newline
Regarding the antichiral edge modes of the mHM, we found that the uniaxial strain could switch their direction of propagation, which may give rise to strain-tuned edge currents. A possible realization of this effect could be achieved in 2D metal transition dichalcogenides, where antichiral edge modes are expected to be observable \cite{Colomes}.
%%%%%%%%%%%%%%%%%%%%%%%%%%%%%%%%%%%%%%%%%%%%%%%%%%%%%%%%%%%%%%%%%%%%%%%%%%%%%%%%%%%%%%%%
\section{Acknowledgment}
 We are indebted to E. Castro and J.-N. Fuchs for stimulating discussions and for a critical reading of the manuscript.

\vspace{1cm}
$^{\ast}$ Electronic address: sonia.haddad@fst.utm.tn

\end{document}